\documentclass{aa}
\usepackage{graphicx}
\usepackage{txfonts}
\usepackage{hyperref}
\newcommand{\refeq}[1]{Eq.~(\ref{#1})}
\newcommand{\reffig}[1]{Fig.~\ref{#1}}

\begin{document} 
\title{Gamma-Ray Burst prompt emission from the synchrotron radiation of relativistic electrons in a rapidly decaying magnetic field}
\titlerunning{GRB prompt emission from the synchrotron radiation in a decaying magnetic field}

   \author{F. Daigne 
          \inst{1,2}
          \and
          \v Z. Bo\v snjak\inst{3,4,1}
          }

   \institute{Sorbonne Université, CNRS, UMR 7095, Institut d’Astrophysique de Paris, 98 bis bd Arago, 75014 Paris, France\\
              \email{daigne@iap.fr}
          \and  
             Institut Universitaire de France
         \and
              Faculty of Electrical Engineering and Computing, University of Zagreb, 10000 Zagreb, Croatia
        \and      
              Laboratoire Univers et Théories, Observatoire de Paris, Université PSL, CNRS, F-92190 Meudon,  France \\
             \email{Zeljka.Bosnjak@fer.hr}
             }

   \date{Received ***; accepted ***}

   \abstract
  {
  Synchrotron radiation from accelerated electrons above the photosphere of a relativistic ejecta
  is a natural candidate for the dominant radiative process
  for the prompt gamma-ray burst emission. 
  There is however a
  tension between the predicted low-energy spectral index $\alpha=-3/2$ in the fast cooling regime and   observations.
  }
  {Radiating 
  electrons have time to travel away
  from their acceleration site,
  and may 
  experience an evolving magnetic field. 
  We study the impact 
  of a decaying 
  field 
  on the synchrotron spectrum.  }
  {We 
  compute
  the radiation from 
  electrons in a decaying magnetic field, 
  including adiabatic cooling, synchrotron radiation, inverse Compton scatterings 
  and pair production. 
  We explore 
  the physical conditions in the comoving frame of the emission region and 
  focus on
  the fast cooling regime where the radiative timescale of electrons with a Lorentz factor $\Gamma_\mathrm{m}$ responsible for the peak of the emission, $t_\mathrm{syn}(\Gamma_\mathrm{m})$, is much shorter than the dynamical timescale $t_\mathrm{dyn}$.
  }
   {We find that the effect of the magnetic field decay 
   depends on 
   its characteristic timescale
   $t_\mathrm{B}$:
   (i) for a slow decay with $t_\mathrm{B}\gtrsim 10\, t_\mathrm{syn}(\Gamma_\mathrm{m})$, the effect is very weak and the spectral shape is mostly determined by
   the impact of the inverse Compton scatterings on the  electron cooling, leading to $-3/2\le \alpha\le -1$; (ii) for a fast decay with $0.1 \,t_\mathrm{syn}(\Gamma_\mathrm{m})\lesssim t_\mathrm{B}\lesssim 10 \,t_\mathrm{syn}(\Gamma_\mathrm{m})$, the magnetic field decay has a strong impact, leading naturally to the synchrotron marginally fast cooling regime, where $\alpha$ tends to $-2/3$ while the radiative efficiency remains high. 
  The high energy inverse Compton component is enhanced in this regime
   ; (iii) for an even faster  decay, the whole electron population 
    is slow cooling.
   }
   {We conclude that efficient synchrotron radiation in a rapidly decaying magnetic field can reproduce low-energy photon indices ranging from $\alpha=-3/2$ to $-2/3$, in agreement with the measured value in the majority of gamma-ray burst 
   spectra. 
    }

   \maketitle{}

\nolinenumbers

\section{Introduction}

The prompt gamma-ray burst (GRB) emission is produced by internal dissipation in a relativistic ejecta \citep{saripiran1997}. However the complete picture that accommodates the observed properties remains elusive largely due to unknowns in the jet composition and the dissipation mechanism. 
Dissipation can start below the photosphere leading to a non-thermal photospheric component \citep[see e.g.][]{rees05,giannios07,
giannios08,beloborodov10,ryde2011,vurm2016,samuelsson22}. Otherwise the photospheric emission is quasi-thermal \citep[see e.g.][]{peer06,peer08,beloborodov10,acuner2020} and remains weak if the initial magnetization is large \citep{daigne02,hascoet13}. In this case, the dominant, non-thermal, component of the prompt GRB spectrum is emitted above the photosphere  in internal shocks \citep{reesmeszaros94,kobayashi97,daigne:98} or magnetic reconnection \citep[see e.g.][]{thompson94,spruit01,drenkhahn02,zhang11,mckinney12}, depending on the magnetization of the ejecta at large distance. 
Synchrotron radiation from accelerated electrons is then a natural expectation, 
but seems in tension with the observed low-energy photon index in prompt GRB spectra  $\alpha$ \citep{band93}, which is usually well above the expected value $-3/2$ in the radiatively efficient regime \citep{sari98}.
This problem was identified in the GRB spectra measured by \textit{CGRO} BATSE \citep{preece98,ghisellini00} and  is confirmed by the
{\it Fermi} Gamma-Ray Burst  Monitor (GBM), that currently provides  broad  spectral information from the hard  X-ray to the soft gamma-ray energy range (8 keV - 40 MeV).  
In  the  latest GBM spectral catalog, 
\cite{poolakkil2021}  reported 
a mean value $\alpha \sim -1.1$ for
the low-energy photon index
resulting from the time-integrated spectral fits and \citet{yu2016} obtain a harder mean value, $\alpha \sim -0.8$ for the time-resolved spectral fits of the brightest bursts observed by {\it Fermi} GBM.

Several processes have been proposed to explain the high values of the observed low-energy photon indices in the context of synchrotron models. In fast cooling regime, inverse Compton (IC) scatterings in the Klein-Nishina regime affect the electron cooling leading to steeper slopes (in $\nu F_\nu$)  of the synchrotron spectrum, reaching  a low-energy photon index $\alpha\sim -1$ 
\citep{derishev01,nakar09,bosnjak09,wang09,daigne11,barniol11}. This leads in addition to a new high-energy spectral component  which may explain the additional component detected by the
{\it Fermi} Large Area Telescope (LAT; $\sim$ 20 MeV to more than 300 GeV)
in the prompt emission of some bright GRBs \citep{ackermann13,ajello19}.
Except in rare cases like GRB\,090926A \citep{ackermann11,yassine17}, this additional prompt high-energy component can usually only be fitted by a power-law, not allowing a detailed comparison to radiative models.

Even higher values of the low-energy photon index in soft gamma-rays, as high as $\alpha=-2/3$, can be measured while maintaining a high radiative efficiency in the regime of marginally fast cooling synchrotron radiation suggested by \citet{daigne11,beniamini:13}. 
Why this radiative regime would be common in GRBs remains however to be understood. In this paper we investigate if this regime could naturally result from the magnetic field evolution from the electron acceleration site to larger scales.

The question of the intensity and structure of the magnetic field in the emission region is complex, and can have a major influence on the synchrotron radiation from accelerated electrons.
Most models for the relativistic ejection by the central engine imply a large scale magnetic field anchored in the source, with a  high initial magnetization \citep[see e.g.][]{spruit01,tchekhovskoy08,tchekhovskoy10,granot11,gottlieb22}. On the other hand, the remaining magnetization at large distance, where the dissipation responsible for the prompt emission occurs, is  debated. In addition, the dissipation process itself will strongly affect the magnetic field at much smaller scales. Shocks, expected if the large scale magnetization at large distance is low, generate a turbulent magnetic field in the shocked region by various plasma instabilities (see e.g. \citealt{sironispitkovskyarons2013}, and \citealt{crumley19,bykov22} in the mildly relativistic regime relevant for  internal shocks). Reconnection, expected when the large scale magnetization at large distance  is high, dissipates the magnetic field and produces a new complex structure in the current sheets, with the formation of magnetic islands \citep[see e.g.][]{drake2006,sironispitkovsky14}. 
Particle-in-cell (PIC) simulations provide 
a self-consistent description of such processes in shocks and reconnection layers, coupled to the particle acceleration
\citep[for recent reviews, see e.g.][]{marcowith20, vanthieghem20}
and magnetic field amplification \citep[see e.g.][]{silva03,medvedev05}.
 However, the size of the simulation box remains limited to the plasma scale (at most a few thousands of plasma skin depth, see e.g. \citealt{keshet09,sironispitkovsky14,crumley19}), which is many orders of magnitude smaller than the geometrical size of the ejecta. The length traveled by radiating electrons responsible for the soft gamma-ray synchrotron component is much longer than this plasma scale, but, is also much smaller than the geometrical size of the ejecta when these electrons are fast cooling. This means that the magnetic field in which these electrons radiate is not described by current simulations, either relativistic MHD simulations of the whole ejecta, or PIC simulations of the dissipative region. 
This paper explores the impact on the observed synchrotron spectrum of a possible evolution of the magnetic field intensity at these intermediate scales, and precisely studies if a magnetic field decay can naturally lead to the marginally fast cooling regime.

Several authors have already suggested models for the prompt and afterglow GRB emission 
where the magnetic field 
decays on lenthscale which is significantly smaller than the geometrical width of the ejecta,
\citet{rossi03,lemoine13} for the afterglow spectrum, and 
\citet{peer06,derishev07,zhao14,zhou:23} 
for the  prompt spectrum.
Following these early studies, we investigate
the effect of such a magnetic field evolution on GRB prompt spectra, 
taking into account in a numerical approach both the synchrotron radiation and the SSC scatterings, and exploring a large parameter space.

In Section~\ref{sec:theory} we discuss
the cooling of electrons in a decaying magnetic field.
Using a numerical radiative code, we study in Section~\ref{sec:results} the additional impact of IC scatterings, and explore in Section~\ref{sec:comparameterspace} the parameter space describing the physical conditions in the comoving frame of the emission region. We also discuss the consequences of the magnetic field decay on the light curves and spectra, as well as on the spectral evolution, in the specific case of internal shocks.
Finally, we discuss in Section~\ref{sec:discussion_conclusion} our results and their comparison to observations, and summarize our conclusions.


\section{Synchrotron radiation in a decaying magnetic field}
\label{sec:theory}

\subsection{Expected effect of a decaying magnetic field}
We assume that electrons are accelerated in a region where the  magnetic field is $B'_0$, corresponding to a magnetic energy density
$u_\mathrm{B,0}=\left.B'_0\right.^2/8\pi$,
and reach a power-law distribution with a minimum Lorentz factor $\Gamma_\mathrm{m}$ and a slope $-p$.
The discussion in this section is entirely done in the comoving frame of the emitting region.
Electrons with Lorentz factor $\gamma$ will radiate their energy on the synchrotron timescale
\begin{equation}
 t'_\mathrm{syn}(\gamma)=t'_\mathrm{dyn}
\frac
{\Gamma_\mathrm{c,0}}{\gamma}\, ,   
\end{equation}
where $\Gamma_\mathrm{c,0}$ is defined as the Lorentz factors of electrons in the magnetic field $B'_0$ that have a synchrotron timescale  equal to the
timescale of the adiabatic cooling (dynamical timescale), i.e.
\begin{equation}
\Gamma_\mathrm{c,0}=\frac{6\pi\, m_\mathrm{e}c}{\sigma_\mathrm{T}\, {B'}^2_0\, t'_\mathrm{dyn}}\, .
\label{eq:defgammac}
\end{equation}
In the fast cooling regime ($\Gamma_\mathrm{m}\gg \Gamma_\mathrm{c,0}$) all electrons radiate efficiently by the synchrotron process. In this case, if the magnetic field is constant ($B'=B'_0$), the peak of the synchrotron spectrum occurs at the synchrotron frequency of electrons at $\Gamma_\mathrm{m}$, i.e. $\nu'_\mathrm{m}=\nu_\mathrm{syn}\left(\Gamma_\mathrm{m}\right)\propto B'_0 \Gamma_\mathrm{m}^2$ and the photon index below the peak is $\alpha=-3/2$ \citep{sari98}. A
photon index
 $-2/3$ 
 (corresponding to a steeper slope $4/3$ in $\nu F_\nu$)
 is recovered at very low frequency below $\nu'_\mathrm{c,0}=\nu_\mathrm{syn}\left(\Gamma_\mathrm{c,0}\right)= \nu'_\mathrm{m}\left( \Gamma_\mathrm{c,0}/\Gamma_\mathrm{m}\right)^{2}\ll \nu'_\mathrm{m}$.
Therefore the \textit{standard fast-cooling} synchrotron spectrum predicts a low-energy photon index $\alpha=-3/2$ which is lower than what is usually observed 
\citep{poolakkil2021}.

A possibility to solve this problem without decreasing the radiative efficiency, which should remain high in gamma-ray bursts to explain the observed luminosities, has been proposed by \citet{daigne11} and \citet{beniamini:13}: in the \textit{marginally fast cooling} regime, where the cooling break $\nu'_\mathrm{c,0}$ is very close to the peak $\nu'_\mathrm{m}$, the intermediate branch of the spectrum with slope $-3/2$ disappears and the large value $\alpha=-2/3$ is recovered. However, maintaining the condition $\Gamma_\mathrm{c,0}\simeq (0.1-1) \Gamma_\mathrm{m} $ during most of the prompt GRB emission may require some kind of fine-tuning of the microphysics parameters.

In this paper, we describe 
how this \textit{marginally fast cooling} regime can naturally emerge if electrons are radiating in a decaying magnetic field. More precisely, we consider the case where
the magnetic field is not constant but rather decays outside  the acceleration site, over a timescale $t'_\mathrm{B}$. 
An electron escaping the acceleration site will radiate in this decaying magnetic field. 
If $t'_\mathrm{B}$ is large compared to $t'_\mathrm{syn}\left(\Gamma_\mathrm{m}\right)=
\frac{\Gamma_\mathrm{c,0}}{\Gamma_\mathrm{m}}
t'_\mathrm{dyn}$, 
electrons with Lorentz factor $\gamma\ga \Gamma_\mathrm{m}$ will still experience a magnetic field of the order of $B'_0$ and the peak and the high-energy part of the synchrotron spectrum will not be affected. On the other hand, if $t'_\mathrm{B}< t'_\mathrm{dyn} $, 
electrons with Lorentz factors $\Gamma_\mathrm{c,0}< \gamma < \Gamma_\mathrm{m}$, will lose their energy more slowly than expected because 
they will meet
 a lower magnetic field when they start to travel outside the initial acceleration site. This will affect the low-energy part of the synchrotron spectrum, as  the cooling break will increase to $\nu_\mathrm{c}\simeq \nu_\mathrm{c,0}\left( t'_\mathrm{dyn}/t'_\mathrm{B}\right)^2$. 
 This allows to naturally tend towards the marginally fast cooling regime. 
Indeed, there are usually several orders of magnitude between the radiative timescale at $\Gamma_\mathrm{m}$ and the dynamical timescale, so that the necessary condition for the proposed scenario to work,
$
t'_\mathrm{syn}
(\Gamma_\mathrm{m}) < t'_\mathrm{B} < t'_\mathrm{dyn}
$,
is not very constraining.
Therefore, a rapidly decaying magnetic field is a promising possibility to recover a  photon index $\alpha \sim -2/3$ in the synchrotron fast cooling regime without reducing the radiative efficiency: it will remain high as long as 
$t'_\mathrm{B}\gg t'_\mathrm{syn}(\Gamma_\mathrm{m})$, i.e. $t'_\mathrm{B}/t'_\mathrm{dyn} \gg \Gamma_\mathrm{c,0}/\Gamma_\mathrm{m}$. This leads to the final condition for the regime described in this paper:
\begin{equation}
\frac{\Gamma_\mathrm{c,0}}{\Gamma_\mathrm{m}} \lesssim \frac{t'_\mathrm{B}}{t'_\mathrm{dyn}} \lesssim 1
\label{eq:conditiondecay}
\end{equation}
Note that if the magnetic field decays extremely fast (${t'_\mathrm{B}}/{t'_\mathrm{dyn}} < {\Gamma_\mathrm{c,0}}/{\Gamma_\mathrm{m}}$), the low-energy photon index $\alpha=-2/3$ is still recovered, but all electrons may be slow cooling, leading to a lower radiative efficiency. On the other hand, if ${t'_\mathrm{B}}/{t'_\mathrm{dyn}} > 1$, the observed spectrum is mostly unaffected by the magnetic field decay.

\subsection{Prescription for the magnetic field decay}

In the following we explore the mechanism described above by assuming a simple prescription for the magnetic field decay:
\begin{equation}
B'(t') = B'_0\, e^{-t'/t'_\mathrm{B}}\, .
\label{eq:evolB}
\end{equation}
\citet{derishev07,lemoine13,panaitescu2019} rather considered a power-law decay of the magnetic field. 
Such a power-law decay is indeed seen in some PIC simulations \citep[see e.g.][]{vanthieghem20}, but at a scale which is much lower than the one considered here.
We tested a power-law prescription, i.e. 
$B'=B'_0$ for $t'\le t'_\mathrm{B}$ and $B'=B'_0\left(t'/t'_\mathrm{B}\right)^{-x}$ for $t'>t'_\mathrm{B}$, with power-law indices $x$ comparable to those used in \citet{lemoineli}, and found that this
does not change significantly the conclusions obtained  when adopting  \refeq{eq:evolB}.
Therefore we only consider \refeq{eq:evolB}  in the following, as it introduces a single new parameter, the timescale $t'_\mathrm{B}$, rather than two parameters in the case of a power-law.

The regime of synchrotron radiation in a rapidly decaying magnetic field
discussed in the present paper 
is similar
to the scenario proposed by \citet{peer06,derishev07,zhao14} for the GRB prompt emission and \citet{rossi03,lemoine13} for the GRB afterglow. In the context of the GRB prompt emission, \citet{peer06} described the effect of a decaying magnetic field on the prompt synchrotron spectrum and gave some orders of magnitude for the expected timescales to show that $t'_\mathrm{B}<t'_\mathrm{syn}$ could indeed be expected in certain conditions in GRBs.
Then \citet{zhao14} presented a first calculation of the radiated spectrum in this scenario in the context of the internal shock model. They used the prescription given by \refeq{eq:evolB}, and also considered a power-law decay,
but made some simplifications: adiabatic cooling is not included, which leads to an over-estimate of the radiative efficiency as it prevents to enter too much in the slow cooling regime, and the Klein-Nishina corrections are not included for the inverse Compton scatterings, whereas they are usually important in the GRB prompt emission phase
\citep{bosnjak09,zou09,piran09}. 
The present paper aims at exploring the same scenario without these simplified assumptions. 
Using the same assumptions than \citet{zhao14}, \citet{zhou:23} studied the additional effect of background magnetic field.
However they focused on a very fast decay,  $t'_\mathrm{B}/t'_\mathrm{dyn}\ll \Gamma_\mathrm{c,0}/\Gamma_\mathrm{m}$. 
As we focus here on less rapid decays (see condition given by Eq.~(\ref{eq:conditiondecay})), the effect of such a background field $B'_\mathrm{bkg}$ should be weak as long as the ratio $f_\mathrm{B}=\left(B'_\mathrm{bkg}/B'_0\right)^2$ is not too large. For this reason, we do not include a background magnetic field in the present study, which avoids introducing an additional free parameter in the model.

Note that \citet{uhm14,geng18,wang2021} also discussed the synchrotron radiation of electrons in a decaying magnetic field in the context of the prompt GRB emission.
However, their scenario is very different from the situation with a \textit{rapid} decay studied here. It corresponds to a \textit{slow} decay of the magnetic field, governed by the large-scale dynamics of the emission zone. Indeed
their prescription is $B'=B'_0 \left(r/r_0\right)^{-b}$ with $b=1.0$
--
$1.5$, and where $r_0$ is the radius where electrons were accelerated (a passive field in a conical jet corresponds to $b=1$, see e.g. \citealt{spruit01}).
This
corresponds to 
a characteristic timescale 
$t'_\mathrm{B}=t'_\mathrm{dyn}/b$
(with $t'_\mathrm{dyn}=\frac{r_0}{\Gamma_0 c}$) that is  much larger than what is considered in the present study. In this scenario, the ratio $t'_\mathrm{B}/t'_\mathrm{dyn}\simeq 0.7-1$
is
very close to the upper limit of the condition given by \refeq{eq:conditiondecay}). 
Therefore, 
the steepest spectra with a photon index $\alpha\to -2/3$ are
obtained by \citet{uhm14} 
by reaching the marginally fast cooling regime with some
fine tuning 
of the initial value $B'_0$ of the magnetic field to get
$\Gamma_\mathrm{c}/\Gamma_\mathrm{m}\sim 0.1-1$.
For instance, the choice of values considered in the reference case
in \citet{uhm14} (see their Fig.~1 and~2) corresponds to $\Gamma_\mathrm{c}/\Gamma_\mathrm{m}\simeq 0.08$.
\citet{geng18} obtained similar results with a calculation extended to include inverse Compton scatterings. In most of their examples, the scatterings in the Klein-Nishina regime have a stronger impact on the spectrum than the slow decay of the magnetic field, leading to a low-energy photon index $\alpha\to 1$, in agreement with \citet{derishev01,nakar09,bosnjak09,daigne11} (see also \S\ref{sec:radiativecode}).

\begin{figure}
    \centering
    \includegraphics[width=\linewidth]{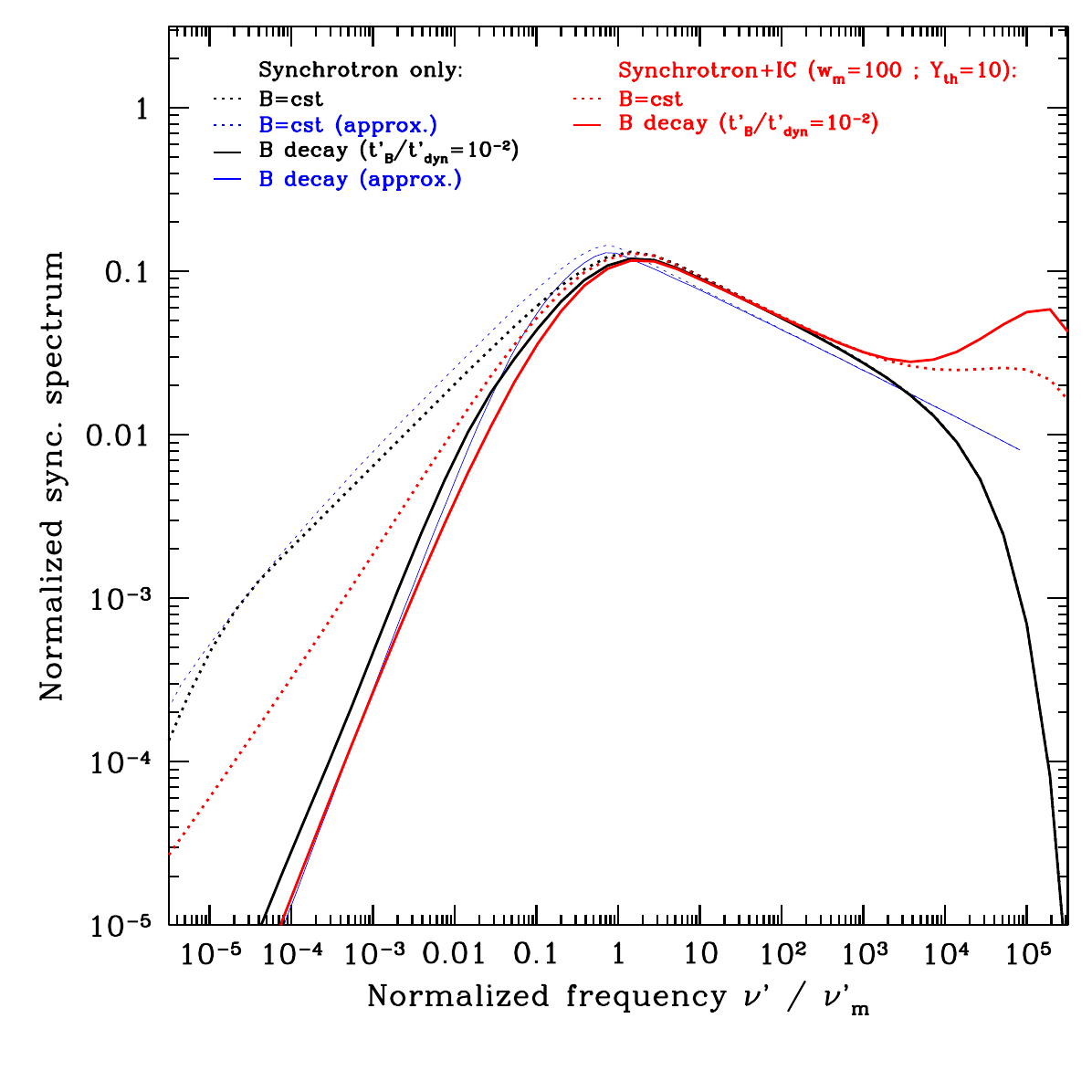}
    \caption{\textbf{Effect of a decaying magnetic field on the synchrotron spectrum.}
    The normalized spectrum $\nu' u_{\nu'}/ u_\mathrm{e}$ is plotted as a function of the normalized frequency $\nu'/\nu'_\mathrm{m}$ for a constant magnetic field (dotted line) or a decaying magnetic field on a timescale $t'_\mathrm{B}=10^{-2} t'_\mathrm{dyn}$ (solid line). The calculation is done with the numerical radiative code described in \S~\ref{sec:radiativecode}, either including only the synchrotron process (black) or both the synchrotron radiation and the inverse Compton scatterings (red). The following parameters are adopted:
$\Gamma_\mathrm{c,0}=\Gamma_\mathrm{m}/300$, $Y_\mathrm{Th}=10$, $w_\mathrm{m}=10^2$, $p=2.5$. In the synchrotron only case, the result of 
the approximate calculation discussed in \S~\ref{sec:semia} is  plotted in blue for comparison.
   }
    \label{fig:example}
\end{figure}

\subsection{Calculation of the synchrotron radiation}
\label{sec:semia}
We assume that the acceleration mechanism proceeds on a timescale much shorter than all other timescales considered here\footnote{This assumption is 
correct for most accelerated electrons, except for the most energetic ones. As we focus here on the peak energy and the low-energy spectral slope, we do not take into account the impact of the magnetic field decay on the maximum Lorentz factor of accelerated electrons, as discussed e.g. in \citet{kumar2012}.
}. Therefore, we adopt a power-law for the initial distribution of electrons at $t'=0$ :
\begin{equation}
    n\left(\gamma,0\right) = (p-1) \frac{n_\mathrm{e}^\mathrm{acc}}{\Gamma_\mathrm{m}}\left(\frac{\gamma}{\Gamma_\mathrm{m}}\right)^{-p}\, ,
\end{equation}
where $n_\mathrm{e}^\mathrm{acc}$ is the density of accelerated electrons in the comoving frame. This corresponds to an initial energy
\begin{equation}
u_\mathrm{e}=\frac{p-1}{p-2}n_\mathrm{e}^\mathrm{acc}\, \Gamma_\mathrm{m}\, m_\mathrm{e}c^2
\end{equation}
 injected in non-thermal electrons. If we neglect all radiative processes except for synchrotron emission, each electron evolves according to
\begin{equation}
    \frac{\mathrm{d}\gamma}{\mathrm{d}t'}=-\frac{\gamma}{t'_\mathrm{dyn}}-\frac{1}{\Gamma_\mathrm{c,0}}\frac{\gamma^2}{t'_\mathrm{dyn}}e^{-2\frac{t'}{t'_\mathrm{B}}}\, ,
\end{equation}
where the first term stands for adiabatic cooling over the timescale $t'_\mathrm{dyn}$ and the second term for synchrotron radiation in a decaying magnetic field, using the prescription given by \refeq{eq:evolB} and the definition of $\Gamma_\mathrm{c,0}$ given by \refeq{eq:defgammac}. The solution for an electron with Lorentz factor $\gamma_0$ is
\begin{equation}
\gamma(t') = \gamma_0  \frac{e^{-t'/t'_\mathrm{dyn}}}{1+\frac{\gamma_0}{\Gamma_\mathrm{c,0}}\frac{1}{1+2\frac{t'_\mathrm{dyn}}{t'_\mathrm{B}}}\left(1-e^{-\left(\frac{1}{t'_\mathrm{dyn}}+\frac{2}{t'_\mathrm{B}}\right)t'}\right)}\, .
\label{eq:solgammae}
\end{equation}
In the standard case (constant magnetic field, $t'_\mathrm{B}\gg t'_\mathrm{dyn}$), the Lorentz factor of a given electron for $t'=t'_\mathrm{dyn}$ equals $\sim \Gamma_\mathrm{c,0}$ for $\gamma_0 \gg \Gamma_\mathrm{c,0}$ (fast cooling electron) and $\sim \gamma_0$ otherwise (slow cooling electron). 
In presence of a decaying magnetic field with $t'_\mathrm{B}\ll t'_\mathrm{dyn}$, it appears clearly from \refeq{eq:solgammae} that $\gamma(t'_\mathrm{dyn})\sim \Gamma_\mathrm{c,0}$ for $\gamma_0 \gg \Gamma_\mathrm{c,0} \frac{t'_\mathrm{dyn}}{t'_\mathrm{B}}$.
Therefore, electrons radiate efficiently only above an effective critical Lorentz factor 
\begin{equation}
    \Gamma_\mathrm{c,eff}\simeq \frac{t'_\mathrm{dyn}}{t'_\mathrm{B}}\Gamma_\mathrm{c,0}\, ,
\end{equation}
which leads to an increase of the cooling break frequency by a factor $\left( {t'_\mathrm{dyn}}/{t'_\mathrm{B}}\right)^2$, as described above. We note that for an extreme decay, i.e. $t'_\mathrm{B}/t'_\mathrm{dyn}\ll \frac{\Gamma_\mathrm{c,0}}{\Gamma_\mathrm{m}}$, we expect a slow cooling spectrum even for $\Gamma_\mathrm{m}>\Gamma_\mathrm{c,0}$, in agreement with Eq.~(\ref{eq:conditiondecay}).

From the solution given by \refeq{eq:solgammae} the electron distribution $n(\gamma,t')=n(\gamma_0,0)\frac{\mathrm{d}\gamma_0}{\mathrm{d}\gamma}$ can be computed at any time $t'$, and the final radiated energy by the synchrotron process can be deduced by integrating the synchrotron power over $t'_\mathrm{dyn}$, i.e.
\begin{equation}
u_\mathrm{syn}=\int_0^{t'_\mathrm{dyn}}\mathrm{d}t'\, \int \mathrm{d}\gamma \, n\left(\gamma,t'\right) \frac{m_\mathrm{e}c^2}{\Gamma_\mathrm{c,0}}\frac{\gamma^2}{t'_\mathrm{dyn}}e^{-2\frac{t'}{t'_\mathrm{B}}}\, .
\end{equation}
The corresponding spectrum $u_{\nu'}$ can also be computed by integrating the synchrotron power at frequency $\nu'$:
\begin{equation}
u_{\nu'}=\int_0^{t'_\mathrm{dyn}}\mathrm{d}t'\, \int \mathrm{d}\gamma \, n\left(\gamma,t'\right) 
P_{\mathrm{syn},\nu'}\left(\gamma; B'(t') \right)
\, .
\end{equation}
This is done in a first step using a simplified shape for the synchrotron spectrum of a single electron with Lorentz factor $\gamma$: a power-law $1/3$ up to $\nu_\mathrm{syn}(\gamma)$ and $0$ above.
The result is plotted in blue in \reffig{fig:example} for $\Gamma_\mathrm{m}/\Gamma_\mathrm{c,0}=300$, either with a constant magnetic field (dotted line) or with a decaying magnetic field with $t'_\mathrm{B}/t'_\mathrm{syn}=10^{-2}$ (solid line), which fulfills the condition
given by \refeq{eq:conditiondecay}.
For the constant magnetic field, we obtain the standard fast cooling synchrotron specrum described by \citet{sari98}, with a photon index $\alpha=-3/2$ below the peak at $\nu'_\mathrm{m}$, and a photon index $-2/3$ below the cooling break $\nu'_\mathrm{c}\simeq 10^{-5}\, \nu'_\mathrm{m}$. For the decaying magnetic field, the effective cooling break is at $\nu'_\mathrm{c}=\nu'_\mathrm{c,0} \left( t'_\mathrm{dyn}/t'_\mathrm{B}\right)^2\sim 10^4\, \nu'_\mathrm{c,0}$, so that the intermediate spectral branch with index $-3/2$ disappears and the photon index $\alpha=-2/3$ is measured immediately below the peak, which is still at $\nu'_\mathrm{m}$.

\begin{figure*}
    \centering
    \begin{tabular}{cc}
    \includegraphics[width=0.45\linewidth]{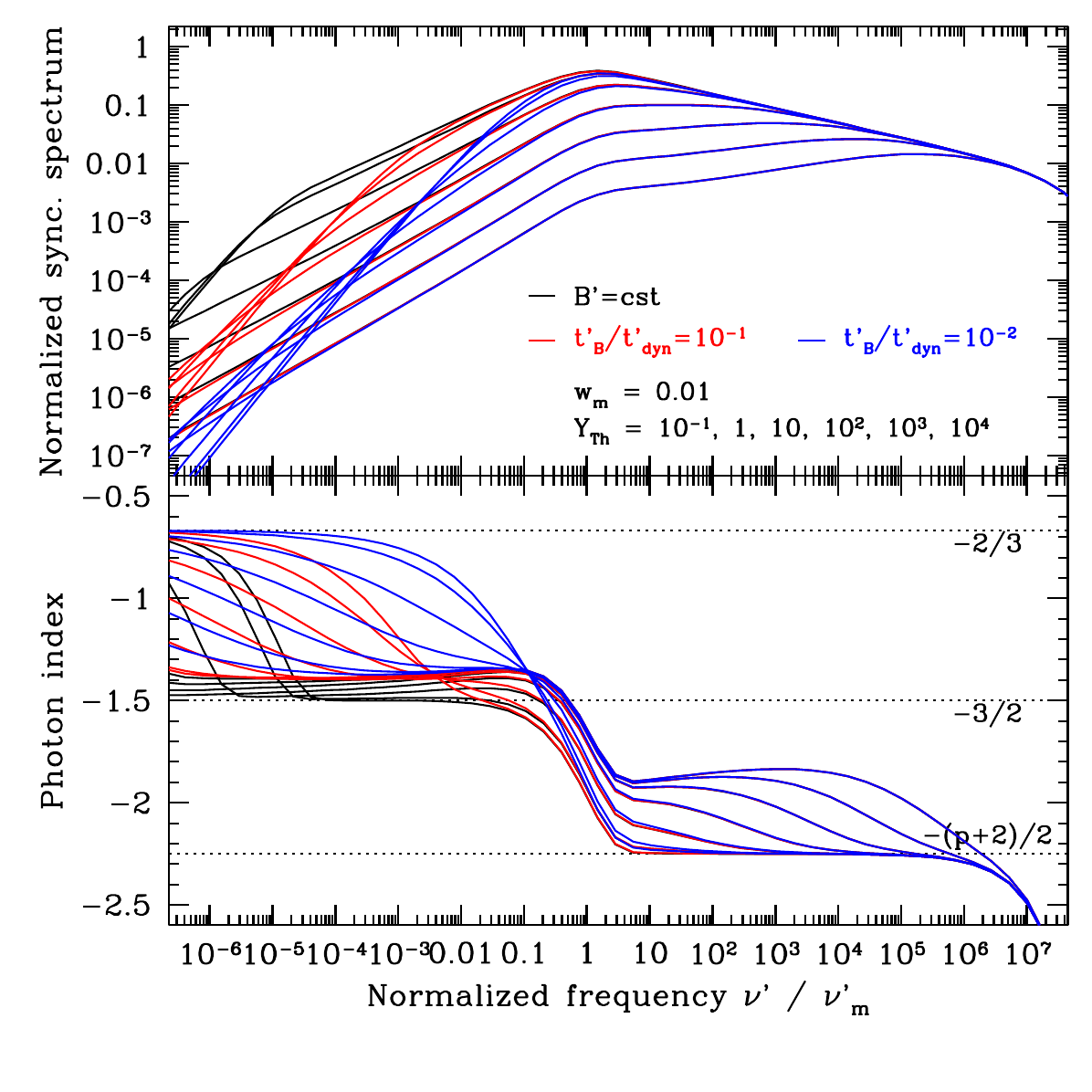} &
    \includegraphics[width=0.45\linewidth]{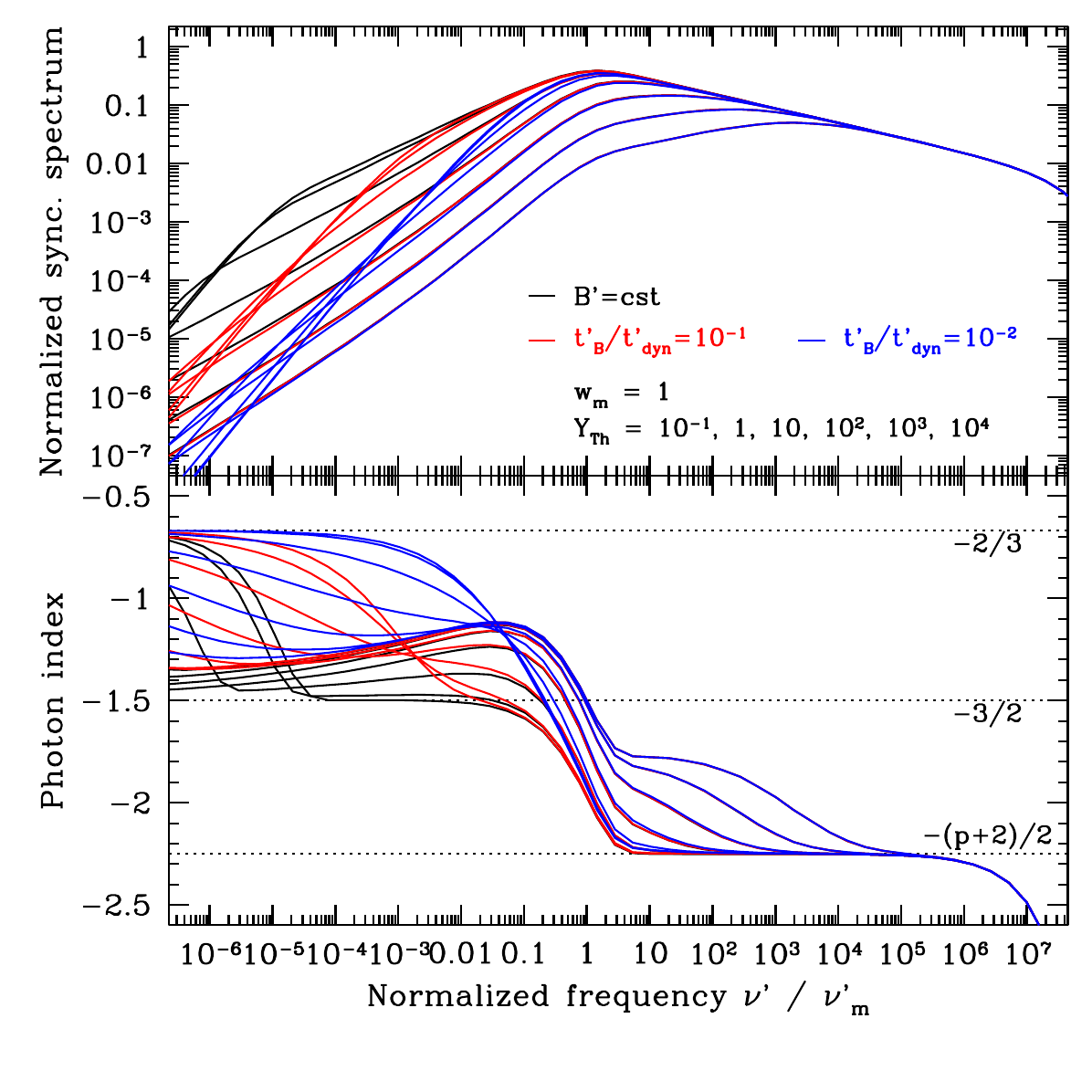}\\
    \includegraphics[width=0.45\linewidth]{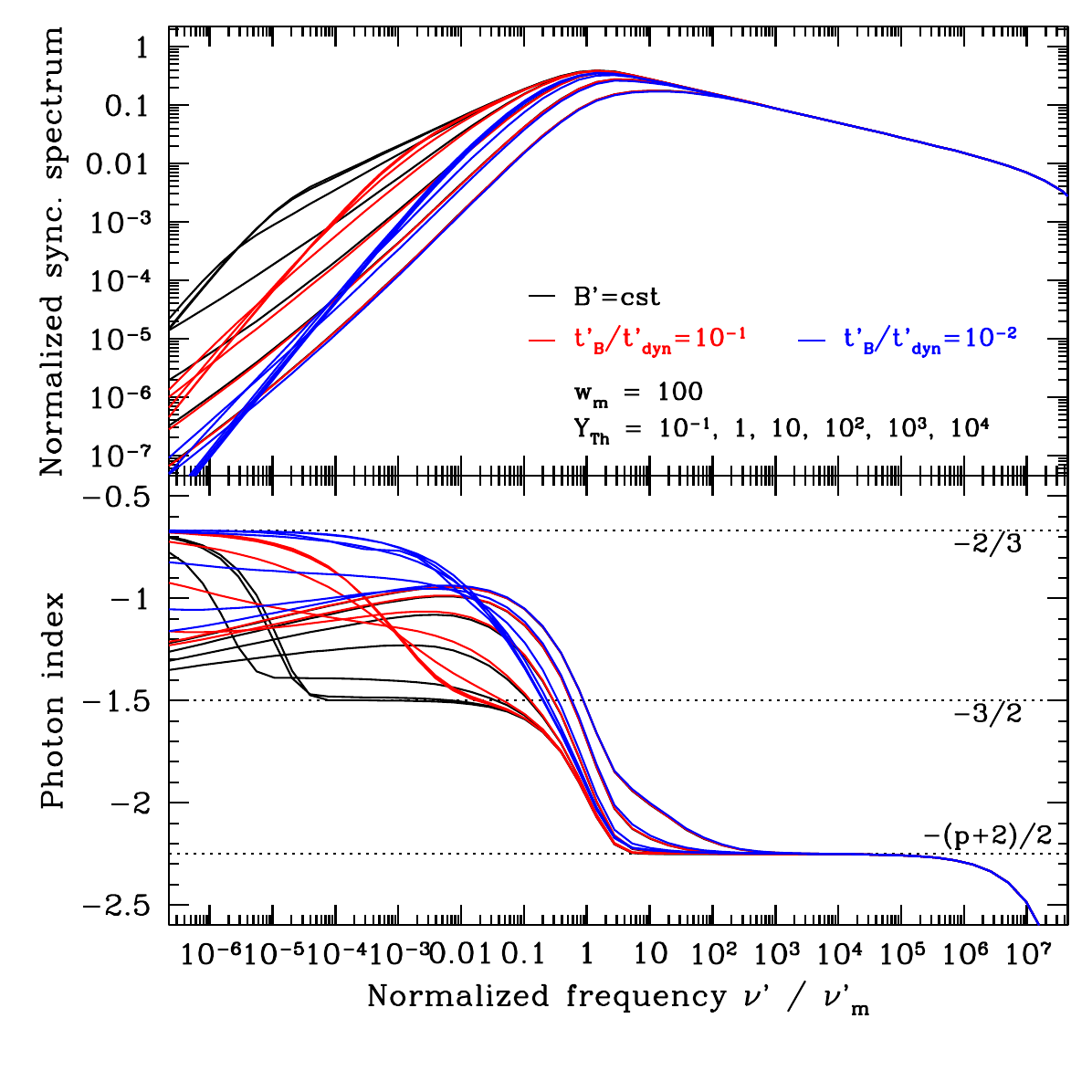} &
    \includegraphics[width=0.45\linewidth]{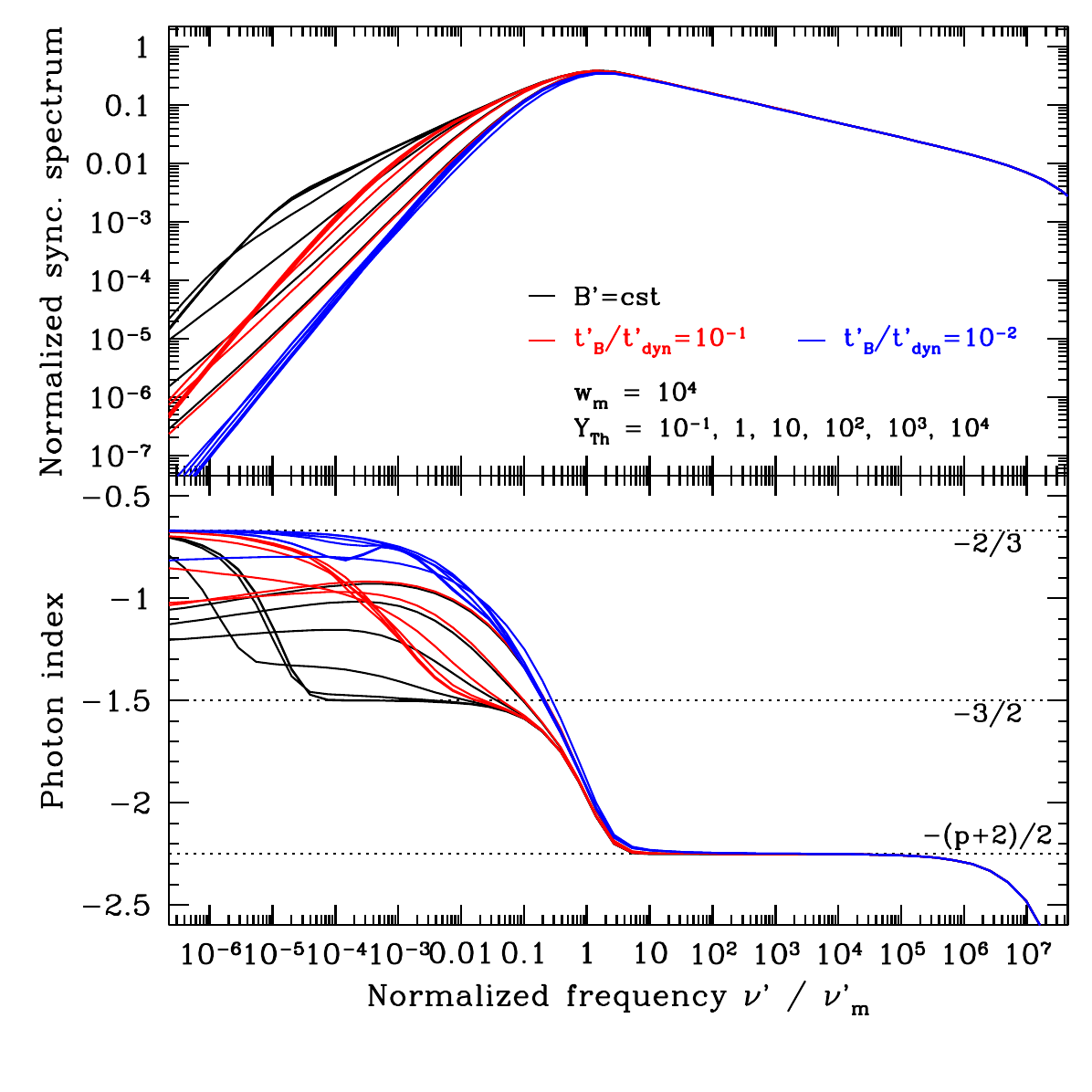}\\
    \end{tabular}
        \caption{\textbf{Effect of a decaying magnetic field on the synchrotron spectrum: realistic case including inverse Compton scatterings.}
    The normalized spectrum $\nu \, u_\nu/ u_\mathrm{e}$, as well as the photon index, are plotted as a function of the normalized frequency $\nu'/\nu'_\mathrm{m}$ for a constant magnetic field (black) or a decaying magnetic field on a timescale $t'_\mathrm{B}/t'_\mathrm{dyn}=10^{-1}
$ (red) and $10^{-2}$ (blue). Each panel correspond to a different value of the parameter $w_\mathrm{m}$ which measures the importance of Klein-Nishina effects, from $10^{-2}$ (Thomson regime) to $10^4$ (strong Klein-Nishina regime). The spectra are plotted for $Y_\mathrm{Th}=10^{-1}$, $1$, $10$, $10^2$, $10^3$ and $10^4$, which corresponds to an increasing efficiency of scatterings (see text).
We assume everywhere an electron slope $p=2.5$ and a magnetic field $B'_0$ in the acceleration site corresponding to $\Gamma_\mathrm{c}/\Gamma_\mathrm{m}=1/300$. 
}
    \label{fig:explore1}
\end{figure*}

\subsection{Additional effects: the role of IC scatterings}
\label{sec:radiativecode}
In a more realistic approach, other radiative processes cannot be neglected. This is especially true for inverse Compton scatterings which can strongly affect the cooling of electrons.
In fast cooling regime, \citet{bosnjak09,daigne11} have shown that  the effect on the synchrotron spectral component is governed by two parameters (see also \citealt{nakar09}):
\begin{equation}
    Y_\mathrm{Th}=\frac{4}{3}\Gamma_\mathrm{m}^2\, \left( \sigma_\mathrm{T}\, n_\mathrm{e}^\mathrm{acc}\, c t'_\mathrm{syn}\left(\Gamma_\mathrm{m}\right)\right)=\frac{p-2}{p-1}\frac{u_\mathrm{e}}{u_\mathrm{B,0}}\, ,
    \label{eq:YTh}
\end{equation}
which governs the efficiency of scatterings,
and
\begin{equation}
    w_\mathrm{m}=\Gamma_\mathrm{m}\frac{h\nu'_\mathrm{m}}{m_\mathrm{e}c^2}\, ,
\end{equation}
which measures how important Klein-Nishina corrections are (Thomson regime corresponds to $w_\mathrm{m}\ll 1$). In particular, \citet{derishev01,nakar09,daigne11} have identified that for $Y_\mathrm{Th}\gg 1$ and $w_\mathrm{m}\gg 1$, 
a photon index $\alpha\lesssim -1$ is obtained below $\nu'_\mathrm{m}$ in the synchrotron component.

Using the radiative code described in \citet{bosnjak09} where a decaying magnetic field following \refeq{eq:evolB} has been implemented, we can solve the evolution of the distribution of electrons, and the corresponding emitted spectrum, in the more realistic case where inverse Compton scatterings are taken into account. This is done for the same example in \reffig{fig:example}, using $Y_\mathrm{Th}=10$ and $w_\mathrm{m}=100$ (red curves). For comparison, the synchrotron only case is also plotted with the same radiative code (black line) and agrees well with the simple calculation presented in \S~\ref{sec:semia}.
When inverse Compton scatterings are included, 
a larger photon index $\alpha\sim -1.2$ is found in the case of a constant magnetic field, due to the effect of scatterings in Klein-Nishina regime, in agreement with \citet{daigne11}. However, an even larger index $\alpha=-2/3$ is obtained when the decay of the magnetic field is included. We note that the inverse Compton scatterings are also modified and we will discuss later the implications for the high-energy prompt emission from gamma-ray bursts. In the following, all results are produced with this same numerical radiative code.

\section{Exploring the effect of the magnetic field decay in different IC regimes}\label{sec:results}

We now explore the parameter space of the synchrotron radiation with a decaying magnetic field. \reffig{fig:explore1} shows the evolution of the synchrotron spectrum as a function of $Y_\mathrm{Th}$ for four different values of $w_\mathrm{m}$ corresponding to different regimes for inverse Compton scatterings, from full Thomson regime to strong Klein-Nishina regime, using either a constant magnetic field (this case is similar to Fig.~2 in \citealt{daigne11}) or a decaying magnetic field with $t'_\mathrm{B}/t'_\mathrm{dyn}=10^{-1}$ or $t'_\mathrm{B}/t'_\mathrm{dyn}=10^{-2}$. 
The initial ratio $\Gamma_\mathrm{c,0}/\Gamma_\mathrm{m}$ is fixed to $1/300$.
In the case of a constant magnetic field (black lines), we recover the results from \citet{daigne11}. In particular, a photon index $\alpha>-1.5$ is found for $w_\mathrm{m}>1$ and $Y_\mathrm{Th}>1$. The limit $\alpha\to-1$ is reached for $w_\mathrm{m}=10^2-10^4$ and $Y_\mathrm{Th}\ga w_\mathrm{m}$. However, as expected from the discussion in Sect.~\ref{sec:theory},
even larger photon indices  are found if the magnetic field is decaying, whatever the values of $w_\mathrm{m}$ and $Y_\mathrm{Th}$ are. Therefore the effect of a magnetic field decay appears as a robust mechanism to produce
a
large photon index (i.e. a steep slope in $\nu F_\nu$)
 in the synchrotron component of the spectrum.

\begin{figure*}
 \centering
   \begin{tabular}{cc}
(a) $Y_\mathrm{Th,0}=0.1$ ; $w_\mathrm{m}=10^{-2}$ 
&
(b) $Y_\mathrm{Th,0}=0.1$ ; $w_\mathrm{m}=10^2$ \\
\includegraphics[height=7.2cm]{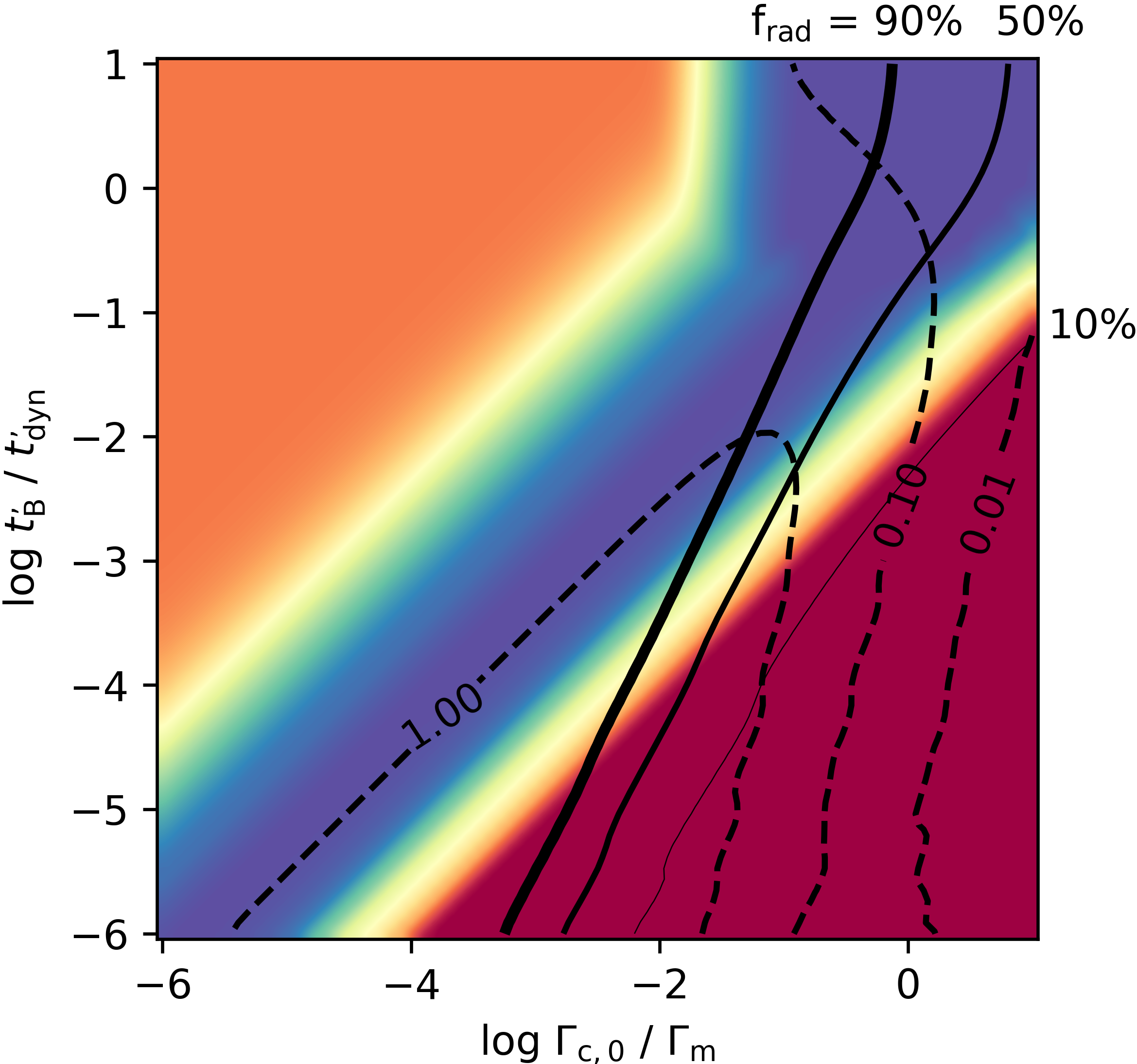} & 
\includegraphics[height=7cm]{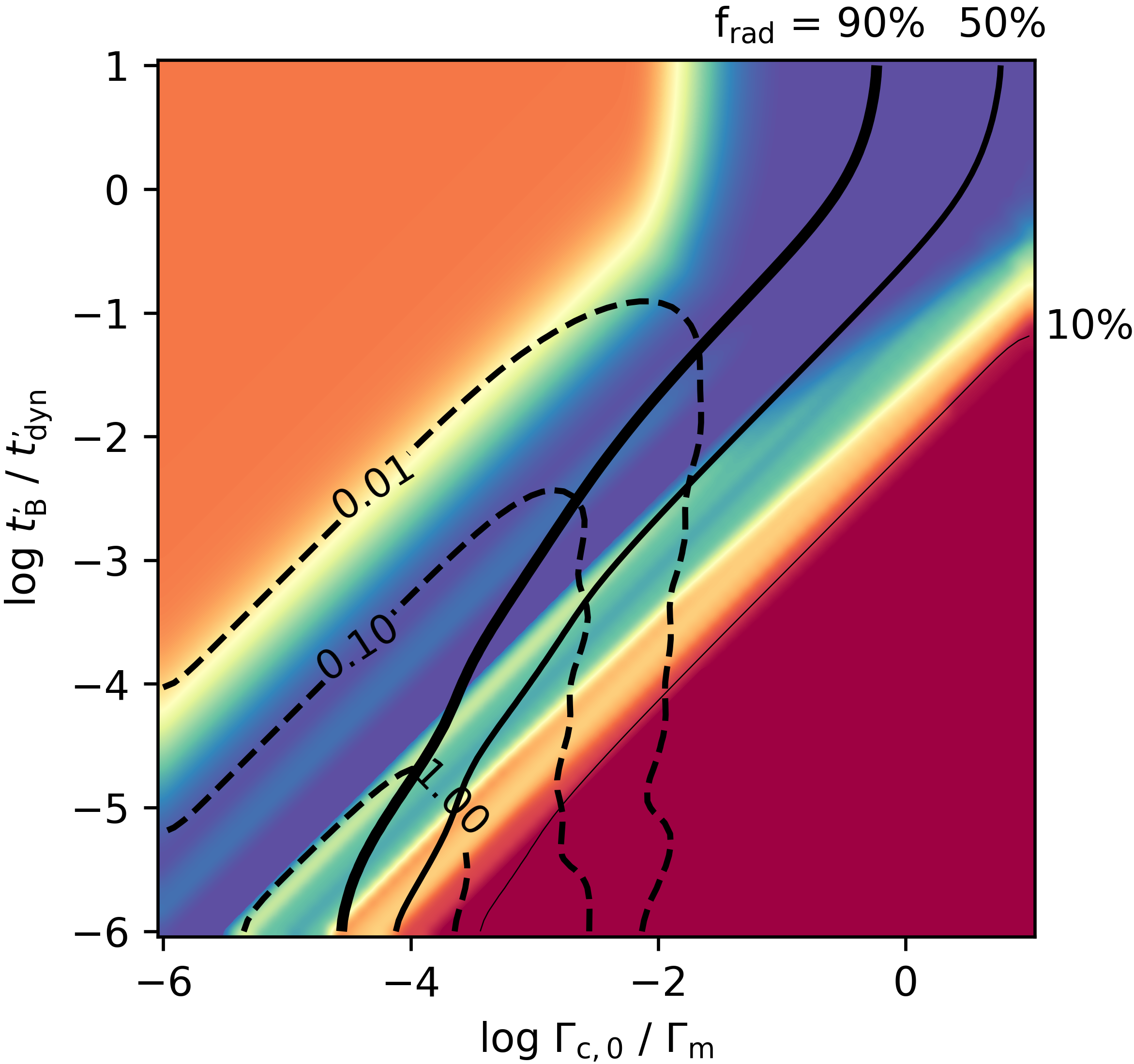}\\
(c) $Y_\mathrm{Th,0}=10^2$ ; $w_\mathrm{m}=10^{2}$ 
& 
(d) $Y_\mathrm{Th,0}=10^2$ ; $w_\mathrm{m}=10^4$ \\
\includegraphics[height=7cm]{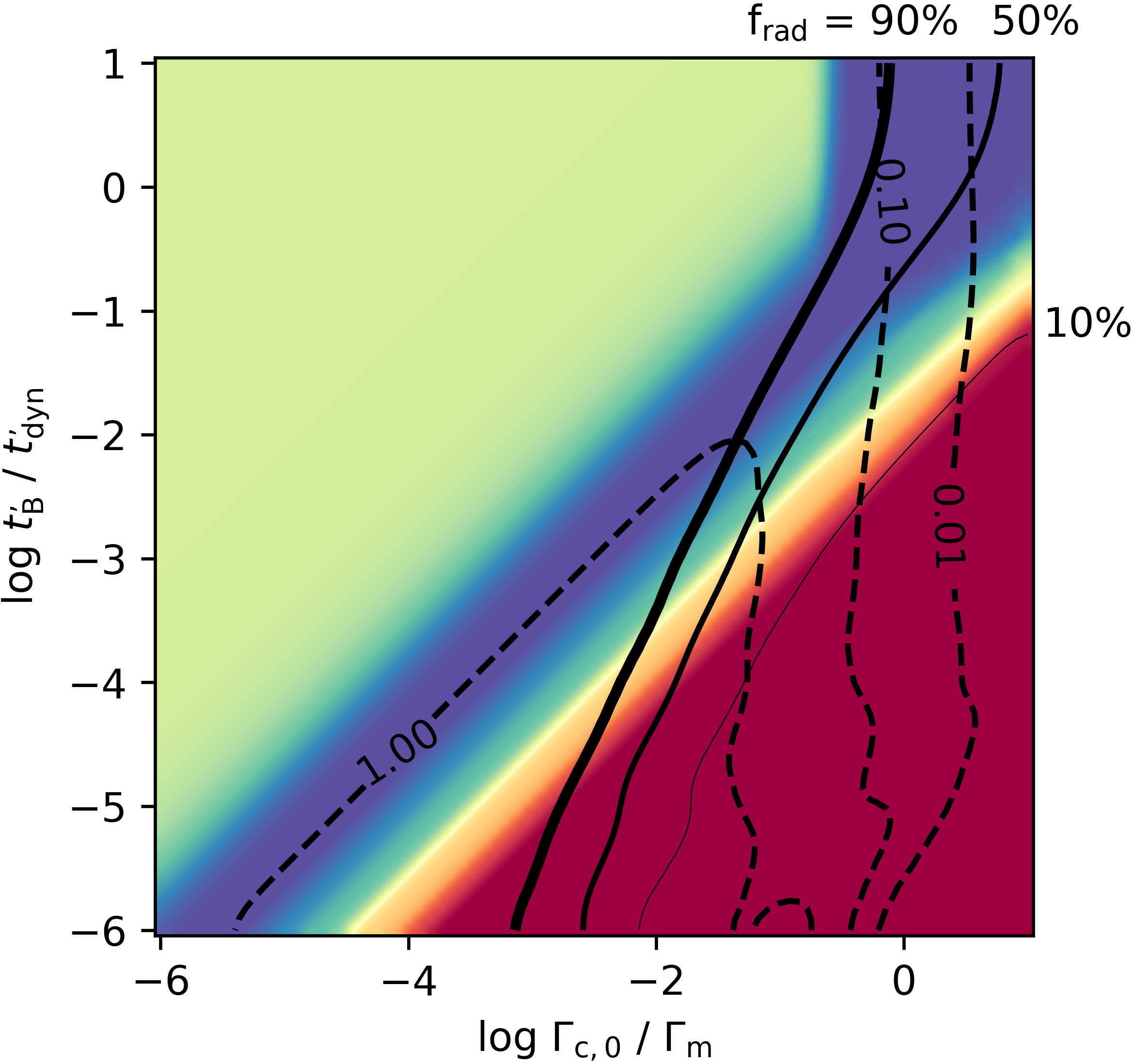} & 
\includegraphics[height=7cm]{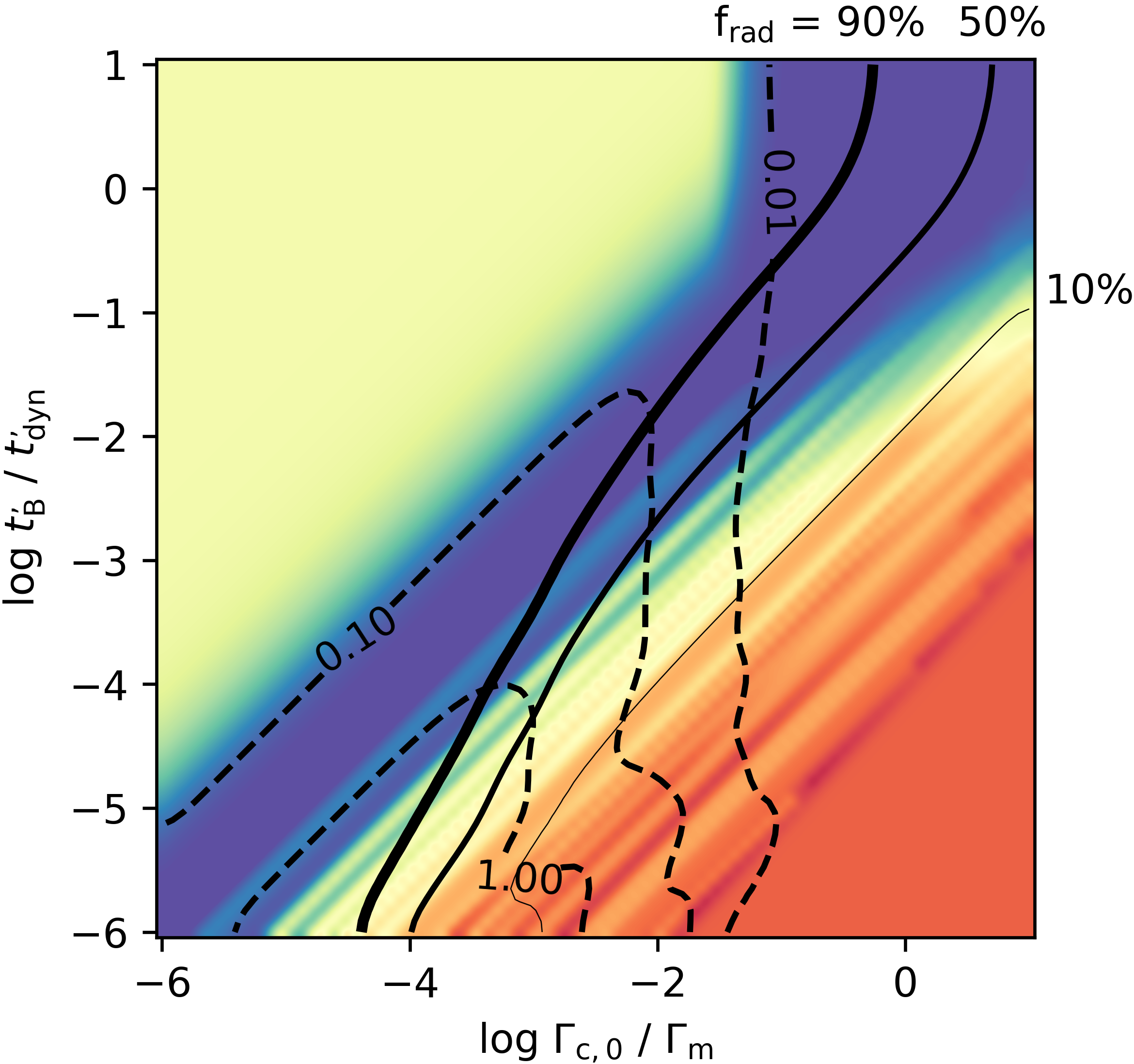}\\ 
    \end{tabular}
    \begin{tabular}{c}
(e) $Y_\mathrm{Th,0}=10^4$ ; $w_\mathrm{m}=10^{4}$\\ 
\includegraphics[height=7cm]{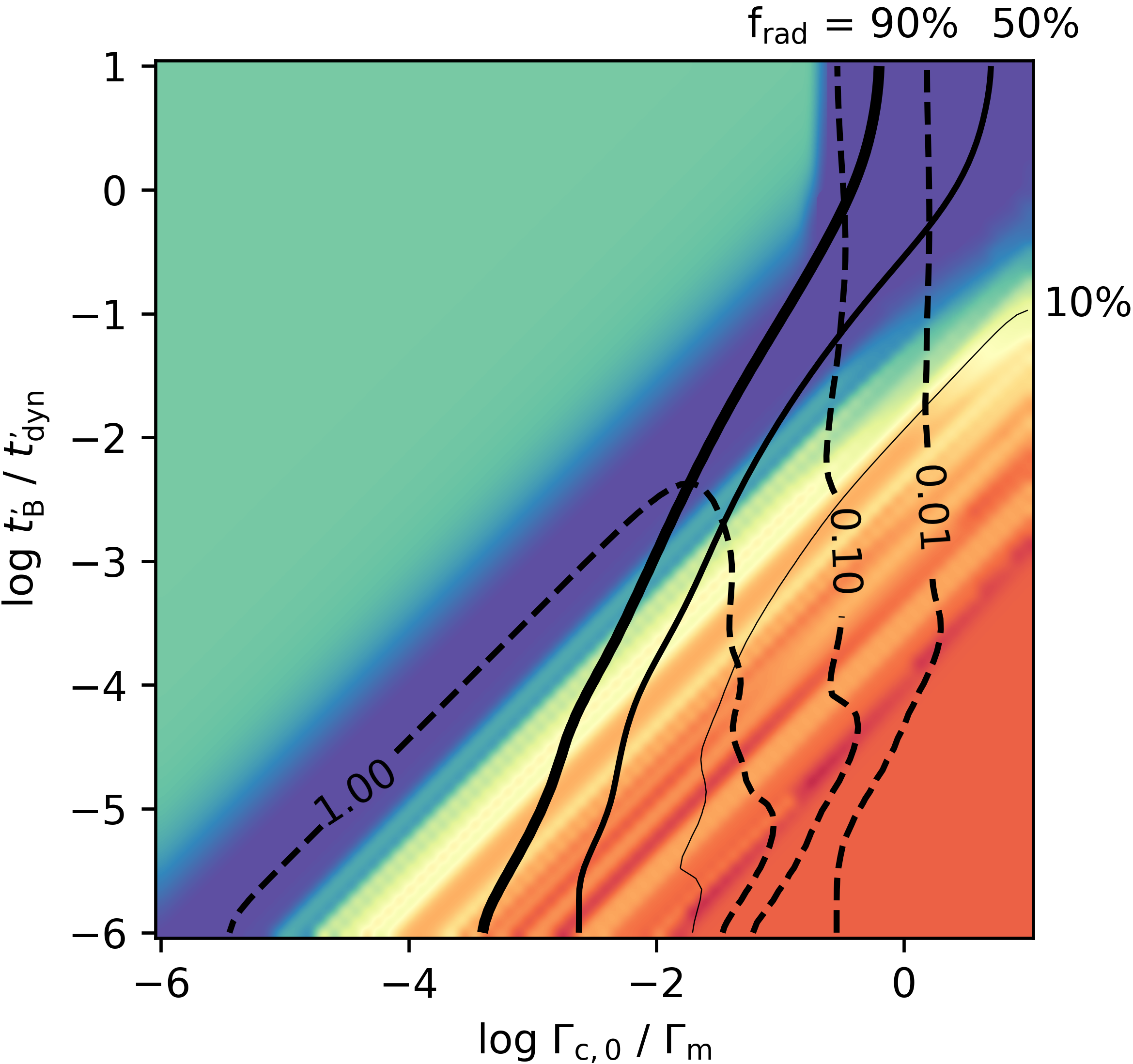} 
\includegraphics[height=7cm]{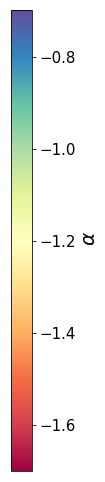}\\
    \end{tabular}
    \caption{\textbf{Synchrotron emission within a decaying magnetic field: parameter space.}
    In the $t'_\mathrm{B}/t'_\mathrm{dyn}$ vs $\Gamma_\mathrm{c,0}/\Gamma_\mathrm{m}$ plane, the value of the low-energy photon index $\alpha$ of the synchrotron spectrum is color-coded.
    In this plane, the standard synchrotron spectrum with a constant magnetic field is at the top ($t'_\mathrm{B}\to\infty$), with the fast cooling regime on the left ($\Gamma_\mathrm{c,0}\ll \Gamma_\mathrm{m}$) and the marginally fast cooling on the right ($\Gamma_\mathrm{c,0}\simeq \Gamma_\mathrm{m}$).
    Black solid lines indicate the limits of the high radiative effiency region (thick: $f_\mathrm{red}=0.9$; thin: $f_\mathrm{dad}=0.5$; very thin: $f_\mathrm{dad}=0.1$). Contours of the inverse Compton/Synchrotron ratio 
    $L_\mathrm{ic}/L_\mathrm{syn}$ are plotted in black dashed lines. Each panel correspond to a set of values for the parameters $Y_\mathrm{Th,0}$ and $w_\mathrm{m}$ that govern the importance of inverse Compton scatterings and of Klein-Nishina corrections.  
}
    \label{fig:diagexplore}
\end{figure*}

\subsection{Photon index and radiative efficiency}
\reffig{fig:diagexplore} shows the value of the photon index below the peak of the synchrotron spectrum in a plane
$t'_\mathrm{B}/t'_\mathrm{dyn}$ versus $\Gamma_\mathrm{c,0}/\Gamma_\mathrm{m}$, for five different sets of the two parameters $\left(Y_\mathrm{Th},w_\mathrm{m}\right)$ governing inverse Compton scatterings.
In this diagram, the standard assumption of a constant magnetic field is recovered at the top, when $t'_\mathrm{B}\gg t'_\mathrm{dyn}$. 
Lines of constant radiative efficiency, 
\begin{equation}
f_\mathrm{rad}=\frac{1}{u_\mathrm{e}}
\int_0^{\infty } u_{\nu'}\, \mathrm{d}\nu'\, ,
\label{eq:radeff}
\end{equation}
are also plotted. 
The gamma-ray burst prompt emission must correspond most of the time to a high radiative effiency to be able to reproduce the observed huge gamma-ray energies and the short timescale variability
\citep{reesmeszaros94,sari96,kobayashi97}. \reffig{fig:diagexplore} shows clearly that in the efficient region ($f_\mathrm{rad}\ga 0.5$) the photon index spans a broad range of values, from the standard fast cooling value $-3/2$ to a maximum value of $-2/3$.
Interestingly, the marginally fast cooling regime $\alpha\simeq -2/3$ is found in a large region of the parameter space.
It can be compared to Fig.~5 in \cite{daigne11}, where this regime was explored for a constant magnetic field and then required some fine tuning of the parameters to maintain a high radiative efficiency ($\Gamma_\mathrm{c,0}/\Gamma_\mathrm{m}\simeq 0.1$--$1$). 

Our numerical calculation shows that the effect of a decaying magnetic field is robust: steep slopes in $\nu F_\nu$ are found for low ratios $\Gamma_\mathrm{c,0}/\Gamma_\mathrm{m}$ and for all values of $Y_\mathrm{Th}$. More precisely:
\begin{itemize}
\item The largest value of the photon index, $\alpha\sim -2/3$ (marginally fast cooling), is obtained 
for a rapid decay with
\begin{equation}
    0.1 \frac{\Gamma_\mathrm{c,0}}{\Gamma_\mathrm{m}}
    \lesssim 
    \frac{t'_\mathrm{B}}{t'_\mathrm{dyn}}
    \lesssim 
    10 \frac{\Gamma_\mathrm{c,0}}{\Gamma_\mathrm{m}}\, ,
\end{equation}
in agreement with Sect.~\ref{sec:theory} (see
\refeq{eq:conditiondecay}).
\item 
Inverse Compton scatterings govern the value of $\alpha$ in the fast cooling regime 
for a slower decay with
$\frac{t'_\mathrm{B}}{t'_\mathrm{dyn}}
    \gtrsim
    10 \frac{\Gamma_\mathrm{c,0}}{\Gamma_\mathrm{m}}$:
when they are negligible (panels (a) and (b)), the standard photon index $\alpha=-3/2$ is recovered; the same value is also obtained when inverse Compton scatterings become important but occur in the Thomson regime (large $Y_\mathrm{Th}$, low $w_\mathrm{m}$, not shown in \reffig{fig:diagexplore}); 
    finally, when scatterings enter the Klein-Nishina regime, $\alpha$ increases towards $-1$, as already discussed in \citet{daigne11}. These results are in agreement with the study by \citet{geng18}, where ${t'_\mathrm{B}}/{t'_\mathrm{dyn}}\sim 1$. 
    \item
    Much flatter $\nu F_\nu$ spectra (photon index $-3/2<\alpha<-2$) are obtained in the bottom-right region of the diagram for a very rapid decay ($ \frac{t'_\mathrm{B}}{t'_\mathrm{dyn}}
    \lesssim 
    0.1 \frac{\Gamma_\mathrm{c}}{\Gamma_\mathrm{m}}$). This is due to the fact that the magnetic field decays so fast that even electron at $\Gamma_\mathrm{m}$ are affected (see corresponding spectra in \reffig{fig:explore1}). 
    This means that the whole electron population enters the slow cooling regime, and the measured low-energy photon index is close to the expected value $-\frac{p+1}{2}$ ($-1.75$ for $p=2.5$) of the intermediate branch between $\nu'_\mathrm{m}$ and $\nu'_\mathrm{c}$ in this case \citep{sari98}.
    As expected the radiative efficiency falls in this region, which cannot correspond to the usual conditions during the GRB prompt emission. 
    Note that a non-negligible background magnetic field can rapidly become dominant in this regime, which would affect the final spectral shape \cite[see][]{zhou:23}.
\end{itemize}

\subsection{The high-energy component}

To investigate the high-energy component of the spectrum, lines of constant ratio
of the inverse Compton luminosity over the synchrotron luminosity 
$L_\mathrm{ic}/L_\mathrm{syn}$ are also plotted in \reffig{fig:diagexplore}.
In agreement with Fermi/LAT observations \citep{ajello19}, 
this figure focus on cases where the inverse Compton component is not dominant (therefore cases with inverse Compton scatterings in Thomson regime (large $Y_\mathrm{Th}$, low $w_\mathrm{m}$) are not shown). Interestingly, the region of steepest synchrotron spectra
shows a large diversity of $L_\mathrm{ic}/L_\mathrm{syn}$ ratios, depending on the regime of inverse Compton scatterings. Largest ratios $L_\mathrm{ic}/L_\mathrm{syn}\simeq 0.1$-$1$ are obtained for large $Y_\mathrm{Th}$ and $w_\mathrm{m} \lesssim Y_\mathrm{Th}$ (\reffig{fig:diagexplore}, panels a, c, e). Much smaller values are obtained if $Y_\mathrm{Th}$ is small ($L_\mathrm{ic}/L_\mathrm{syn}\simeq 10
^{-2}$-$10^{-1}$ in panel~b) or if the Klein-Nishina reduction becomes strong ($L_\mathrm{ic}/L_\mathrm{syn}\simeq 10
^{-3}$-$10^{-2}$ in panel~d). This may explain the observed diversity revealed by the Fermi satellite: during the prompt phase, the ratio of the  fluence at high energy (0.1-100 GeV) measured by the LAT instrument over the fluence at low energy (10 keV-1 MeV) measured by the GBM instrument 
is typically in the range $10^{-2}-1$ in GRBs detected by the LAT  (see Fig.~15 in  the second Fermi-LAT GRB catalog, \citealt{ajello19}), with probably even lower ratios for GBM bursts that are not detected by the LAT. 

Current observations at very high energies ($>$ 100 GeV) were attributed to the afterglow phase \citep{magicgrb, 2019Natur.575..464A,lhaaso}.
 The future facilities, e.g. Cherenkov Telescope Array (CTA; \citealt{ctabook}) 
 may open 
 new perspectives to study the possible Synchrotron Self-Compton (SSC) component in prompt GRB spectra and impose better constraints on the parameters.  


\section{Exploring the physical conditions in the comoving frame of the emission region}
\label{sec:comparameterspace}

\subsection{Impact of each physical parameter}

\begin{figure*}
    \centering
    \begin{tabular}{cc}
    \includegraphics[width=0.45\linewidth]{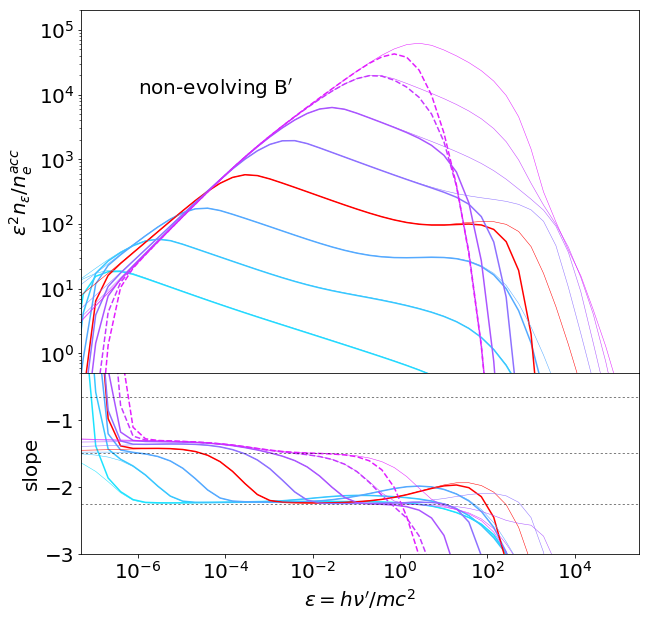} & 
    \includegraphics[width=0.45\linewidth]{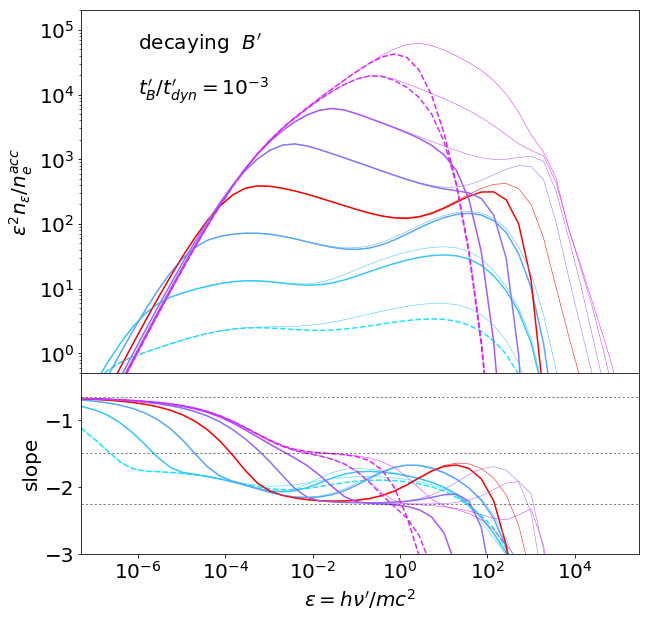}\\
    \end{tabular}
     \caption{\textbf{Emission in the comoving frame: effect of the electron Lorentz factor.} 
     Starting from the reference case shown in red ($\Gamma_\mathrm{m}$ = 1600, B$'_0$ = 2000 G, n$_\mathrm{e,acc}$ = 4.1$\times$10$^7$ cm$^{-3}$, t$_{\mathrm{dyn}}$ = 80 s), we plot the evolution of the spectrum when varying the electron Lorentz factor, either assuming a constant  (left) or a decaying (right) magnetic field. The adopted values are $\Gamma_{\mathrm{m}}$ = 51 (cyan), 160, 510, 1600, 5100, 1.6$\times$10$^4$, 5$\times$10$^4$, 1.6$\times$10$^5$ (magenta).
     The spectra obtained when pair production and  synchrotron self-absorption are not included 
     are shown in thin lines. Dashed lines show the spectra that do not satisfy the conditions for transparency ($\tau_\mathrm{T} <$ 0.1), or 
     are radiatively inefficient ($f_\mathrm{rad}<50\%$).
     }
    \label{fig:explore4_gmin}
\end{figure*}

\begin{figure*}
    \centering
    \begin{tabular}{cc}
    \includegraphics[width=0.45\linewidth]{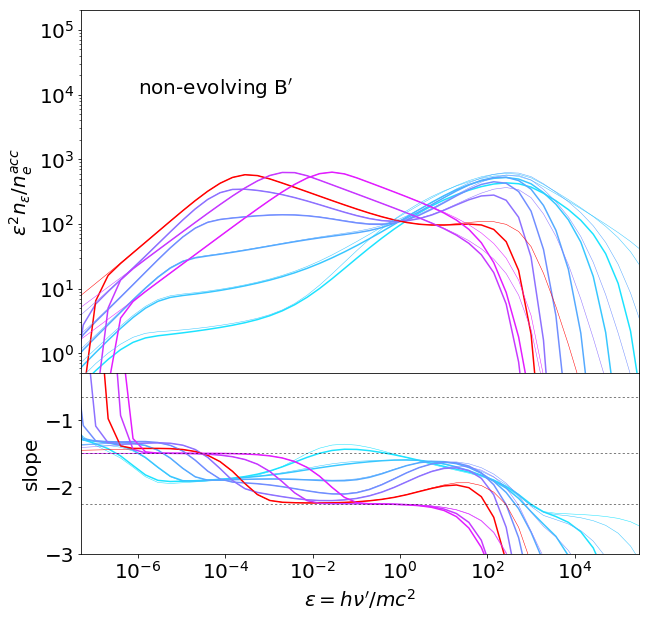} & 
    \includegraphics[width=0.45\linewidth]{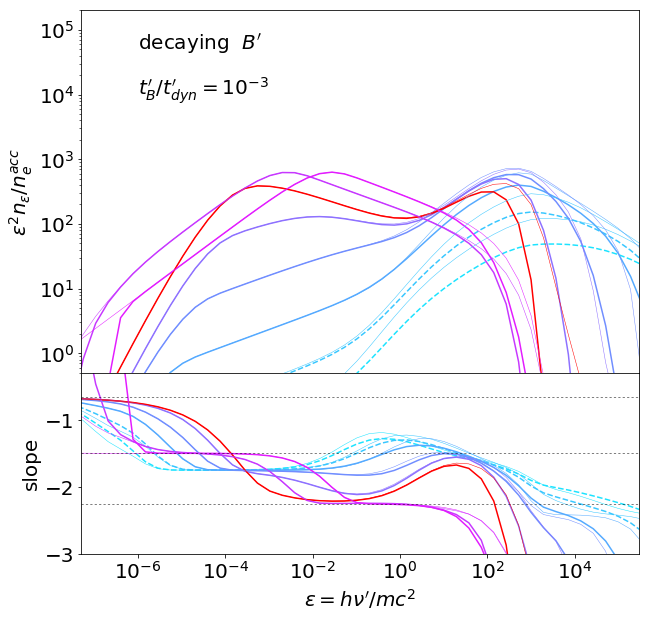}\\
    \end{tabular}
     \caption{\textbf{Emission in the comoving frame: effect of the magnetic field.} 
     Same as in Fig.\ref{fig:explore4_gmin}, now varying the initial magnetic field. The adopted values are
      B$'_0$ [G] = 6 (cyan), 20, 63, 200, 632, 2000, 2$\times$10$^4$, 2$\times$10$^5$ (magenta).
      }
    \label{fig:explore4_B}
\end{figure*}

\begin{figure*}
    \centering
    \begin{tabular}{cc}   
       \includegraphics[width=0.45\linewidth]{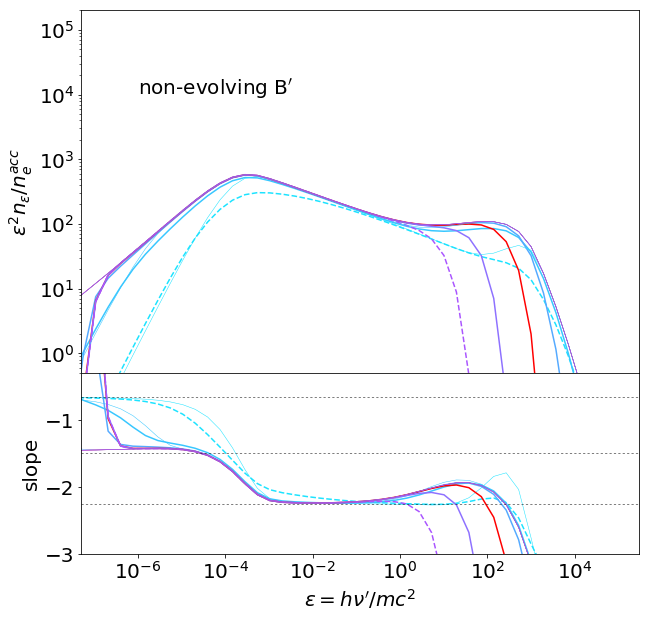} & 
    \includegraphics[width=0.45\linewidth]{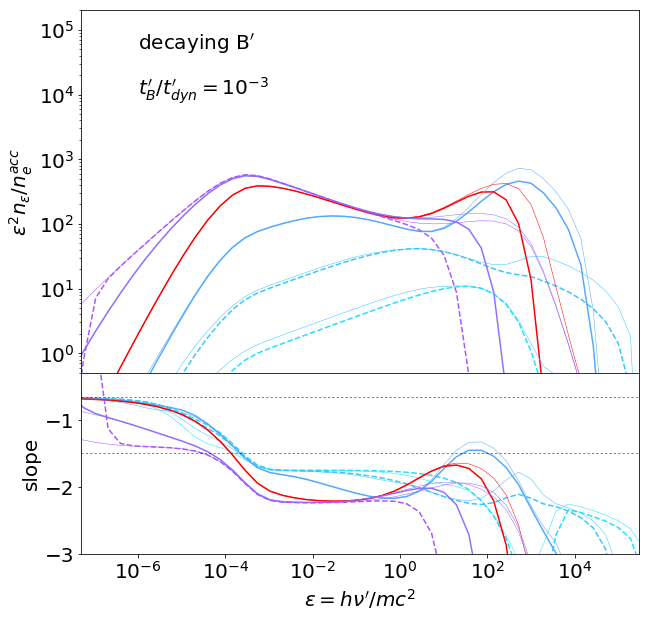}\\
    \end{tabular}
        \caption{\textbf{Emission in the comoving frame: effect of the dynamical timescale.} 
        Same as in Fig.\ref{fig:explore4_gmin}, now varying the dynamical timescale. The adopted values are
       t$_\mathrm{dyn}$ [s] = 0.08 (cyan), 0.8, 8, 80, 800, 8$\times$10$^3$ (magenta). 
      }
    \label{fig:explore4_tdyn}
\end{figure*}

\begin{figure*}
    \centering
    \begin{tabular}{cc}   
    \includegraphics[width=0.45\linewidth]{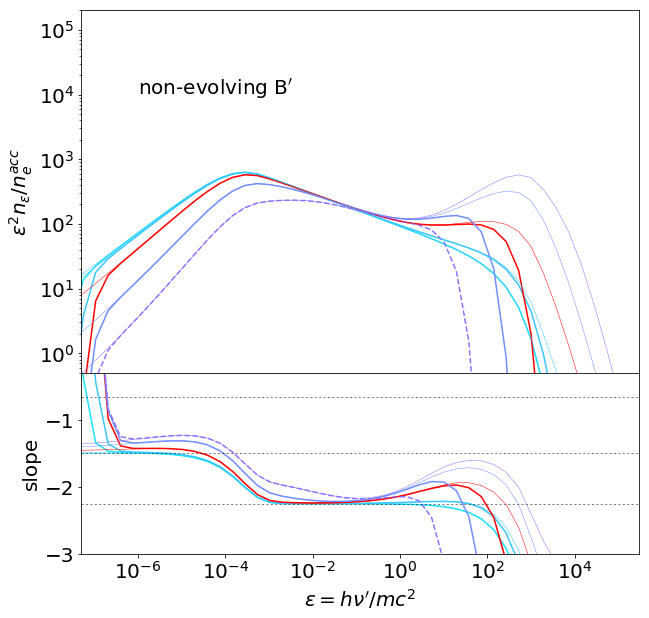} &
    \includegraphics[width=0.45\linewidth]{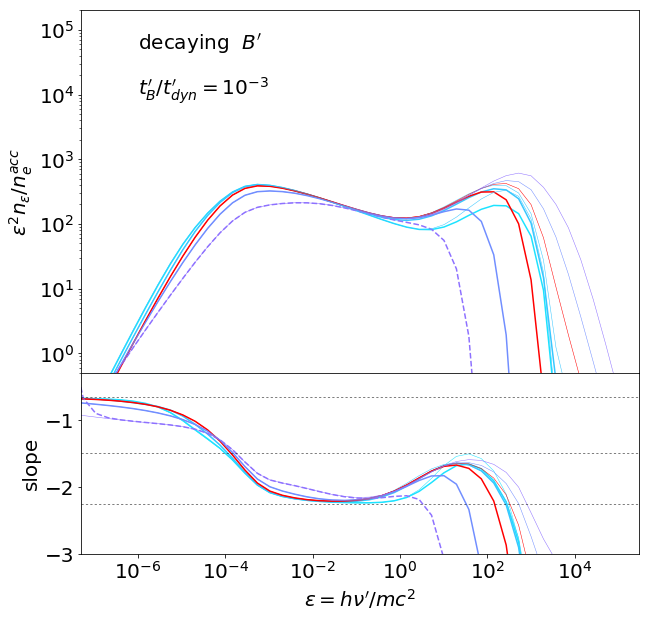}\\
 
    \end{tabular}
        \caption{\textbf{Emission in the comoving frame: effect of the electron density.} 
        Same as in Fig.\ref{fig:explore4_gmin}, now varying the electron density. The adopted values are
      n$_\mathrm{e,acc}$ [cm$^{-3}$]= 4.1 $\times$ 10$^5$ (cyan), 4.1 $\times$ 10$^6$ ,4.1 $\times$ 10$^7$ ,4.1 $\times$ 10$^8$ ,4.1 $\times$ 10$^9$ (magenta). 
        }
    \label{fig:explore4_ne}
\end{figure*}

The radiated spectrum in the comoving frame of the emitting region is entirely determined by five parameters already discussed in \citet{bosnjak09}, namely the magnetic field $B'_0$, the adiabatic cooling time scale $t'_{\mathrm{dyn}}$, the density of accelerated electrons  $n_{\mathrm{e}}^{\mathrm{acc}}$,  the minimum Lorentz factor of the accelerated electrons $\Gamma_{\mathrm{m}}$ and the slope $-p$ of their power-law distribution, to which must be added the timescale $t'_{\mathrm{B}}$ when a possible magnetic field decay is considered. 
We adopt the same reference case for these quantities as in \citet{bosnjak09}, $B'_0 = 2000\, \mathrm{G}$, $t'_{\mathrm{dyn}} = 80\, \mathrm{s}$, $n_\mathrm{e}^{\mathrm{acc}} = 4.1 \times 10^{7}\, \mathrm{cm^{-3}}$,  $\Gamma_{\mathrm{m}} = 1600$, and $p=2.5$ (see their Fig.~6).
This leads to $\Gamma_\mathrm{c,0}/\Gamma_\mathrm{m}=1.5\times 10^{-3}$ (fast cooling).
We then study the effect on the spectrum of each physical parameter by varying its value while keeping all other parameters constant. 
The results are shown in \reffig{fig:explore4_gmin} (effect of $\gamma_\mathrm{m}$), \reffig{fig:explore4_B} (effect of $B'_0$), \reffig{fig:explore4_tdyn} (effect of $t'_\mathrm{dyn}$) and
\reffig{fig:explore4_ne} (effect of $n_\mathrm{e}^\mathrm{acc}$), either for a constant magnetic field (left panels in the figures) or for a decaying magnetic field with $t'_\mathrm{B}/t'_\mathrm{dyn}=10^{-3}$.
The panels with a constant magnetic field  just reproduce the results in Fig.~6 of \citet{bosnjak09} for an easier comparison with the decaying case  discussed here. As expected,
the magnetic field decay leads to a clear increase of 
 the low-energy photon index of the synchrotron component in most cases. We detail below the effect of each physical parameter.

\paragraph{Effect of the electron minimum Lorentz factor (\reffig{fig:explore4_gmin}).} 
The peak energy of the synchrotron spectrum increases with $\Gamma_\mathrm{m}$. Compared to the standard case, the magnetic field decay has two effects: (i) as expected the low-energy photon index increases, due to a larger effective critical Lorentz factor,  $\Gamma_\mathrm{c,eff}/\Gamma_\mathrm{c,0}\simeq t'_\mathrm{dyn}/t'_\mathrm{B}=10^3$ (see the steepening of the $\nu u_\nu$ spectrum at low frequency in the top-right panel); (ii) in addition, the intensity of the IC component is increased. This is due to a longer synchrotron timescale. The effect is stronger for low values of $\Gamma_\mathrm{m}$ as the KN regime strongly limit the IC scatterings for higher synchrotron peak energies. Dashed lines indicate the limits of the regime of interest: the radiative efficiency becomes too low for small electron Lorentz factor (slow cooling), and the optically thin assumption is not valid anymore for very large electron Lorentz factor, due to the production of pairs by high-energy photons.

\paragraph{Effect of the magnetic field (\reffig{fig:explore4_B}).} 
When $B'_0$ decreases, the efficiency of IC scatterings increases (lower Compton parameter, Eq.~(\ref{eq:YTh})) but the peak energy of the synchrotron spectrum also decreases, so that the lowest values of $B'_0$ correspond to a low radiative efficiency (slow cooling, dashed lines). The IC component is dominant in this case (Thomson regime), which is unlikely in GRB spectra \citep{bosnjak09,piran09}. For higher values of the magnetic field, the KN regime limits the intensity of the IC component at high-energy but contributes to the steepening of the low-energy synchrotron slope  in the $\nu\, u_\nu$ spectrum \citep{nakar09,bosnjak09,daigne11}. Compared to the standard case, the magnetic field decay has two effects: (i) the expected increase of the low-energy photon index in the synchrotron component; (ii) an increased intensity of the high-energy IC component, due to lower effective magnetic field energy density.   
\paragraph{Effect of the dynamical timescale (\reffig{fig:explore4_tdyn}).}
The dynamical timescale directly impacts the critical Lorentz factor $\Gamma_\mathrm{c,0}$ (Eq.~(\ref{eq:defgammac})), and therefore the radiative regime of electrons.
To avoid a low radiative efficiency, electrons should be in fast cooling, which constraints  $t'_\mathrm{dyn}$ to be greater than the synchrotron timescale $t'_\mathrm{syn}(\Gamma_\mathrm{m})$, equal to $0.12\, \mathrm{s}$ in the reference case. The lowest values of $t'_\mathrm{dyn}$ in
the left panel of \reffig{fig:explore4_tdyn} 
therefore correspond to the marginally fast cooling, with an increase of the low-energy photon index. Compared to the standard case, the magnetic field decay leads again to the two expected effects: (i) the marginally fast cooling regime is more easily reached, with a clear increase of the low-energy photon index. This effect disappears for the largest values of $t'_\mathrm{dyn}$ as $\Gamma'_\mathrm{c,0}$ becomes very small; (ii) the intensity of the IC component at high energy is enhanced. The regime of interest remains limited by the condition on the radiative efficiency ($f_\mathrm{rad}\gtrsim 50\%$ for $t'_\mathrm{dyn}\gtrsim 1\, \mathrm{s}$) and by the increasing opacity due to secondary pairs (optically thick regime for $t'_\mathrm{dyn}\gtrsim 8000\, \mathrm{s}$). 

\paragraph{Effect of the density of accelerated electrons 
(\reffig{fig:explore4_ne}).}
The intensity of the IC component at high energy increases with the electron density (Eq.~(\ref{eq:YTh})) but it saturates for  the highest densities  because of the pair production. The magnetic field decay has a similar effect at all densities, with a clear increase of the low-energy photon index of the synchrotron component. 

\subsection{Consequences for the spectral evolution during the prompt GRB emission}
\label{sec:pulse_is}
During the prompt GRB phase, there are several emission zones in the relativistic ejecta, with possibly different physical conditions (which may evolve).
Therefore some spectral evolution is expected, in agreement with observations.
In particular, studies
 of bright, isolated pulses have demonstrated that there are typical patterns occurring in prompt GRB spectral evolution, e.g. the hard-to-soft evolution of peak energy $E_\mathrm{p}$ with time, the correlation between intensity and the peak energy $E_\mathrm{p}$, and the low-energy photon index $\alpha$ evolving with $E_\mathrm{p}$ \citep[see e.g.][]{ford95, crider97, peng09, lu12, hakkila15, yu2016}.

In the context of synchrotron radiation with a decaying magnetic field, we expect from the discussion in the previous subsection that both the peak energy and the low-energy spectral slope should evolve together with the physical properties in the emission region.
In order to model this 
spectral evolution,  
a specific internal dissipation process must be assumed in the relativistic ejecta.
This is done here in the context of the internal shock model, following the approach described in \citet{bosnjak09}.

\subsubsection{Evolution of the physical conditions in internal shocks, and associated spectral evolution}

We model a typical pulse in a GRB gamma-ray lightcurve produced by the emission of electrons accelerated in internal shocks resulting from the collision of a fast region with a Lorentz factor $\Gamma_2$ and a slow region with a Lorentz factor $\Gamma_1 < \Gamma_2$. We study the same references cases as in \citet{daigne11,bosnjak14}: we consider an ejection lasting $t_\mathrm{w}=2\, \mathrm{s}$ with a constant injected kinetic power $\dot{E}=10^{54}\, \mathrm{erg/s}$ and a Lorentz factor evolving from $\Gamma_1=100$ to $\Gamma_2=400$. We assume that a fraction $\epsilon_\mathrm{e}$ of the energy dissipated in internal shocks is injected into non-thermal electrons that represent a fraction $\zeta$ of all electrons, and that a fraction $\epsilon_\mathrm{B}$ of the dissipated energy is injected in an amplified magnetic field of intensity $B'_0$ in the acceleration site. The index of the power-law distribution of accelerated electrons is $p=2.5$. Case (A) assumes $\zeta=3\times10^{-3}$ and $\epsilon_\mathrm{B}=1/3$ (high magnetic field, very weak inverse Compton emission) and case (B) $\zeta=10^{-3}$ and $\epsilon_\mathrm{B}=10^{-3}$ (low magnetic field). In both cases, the synchrotron component is dominant and peaks in the gamma-ray range. Without any magnetic field decay, case (A) corresponds to the standard fast-cooling synchrotron spectrum with 
$E_\mathrm{p}\simeq 700\, \mathrm{keV}$ and $\alpha=-3/2$ and case (B) to a harder spectrum with 
$E_\mathrm{p}\simeq 800\, \mathrm{keV}$ and $\alpha\simeq -1$, due to the effect of IC scatterings in Klein-Nishina regime. 

\begin{figure*}
    \centering
    \begin{tabular}{cc}
    \includegraphics[width=0.48\linewidth]{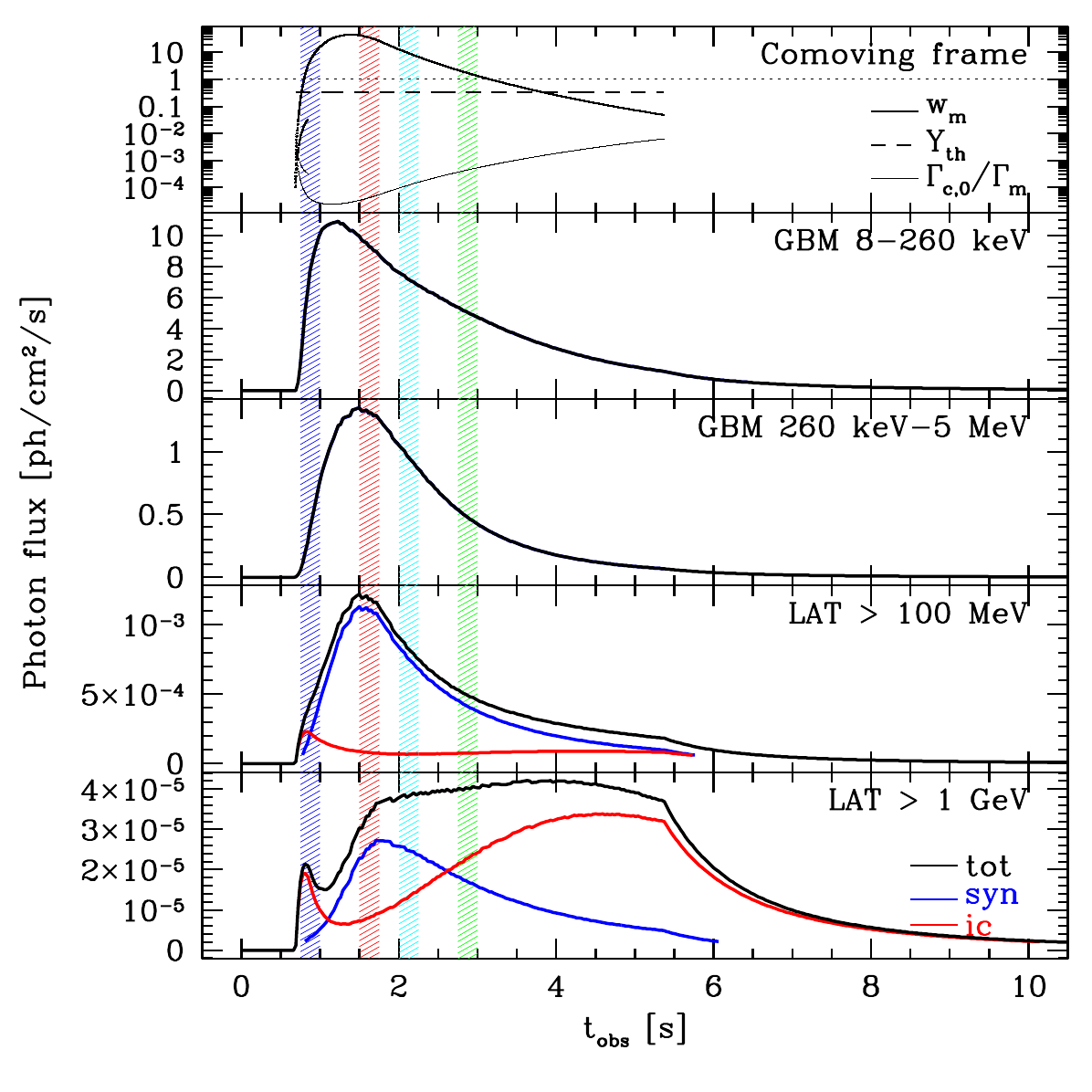} & 
    \includegraphics[width=0.48\linewidth]{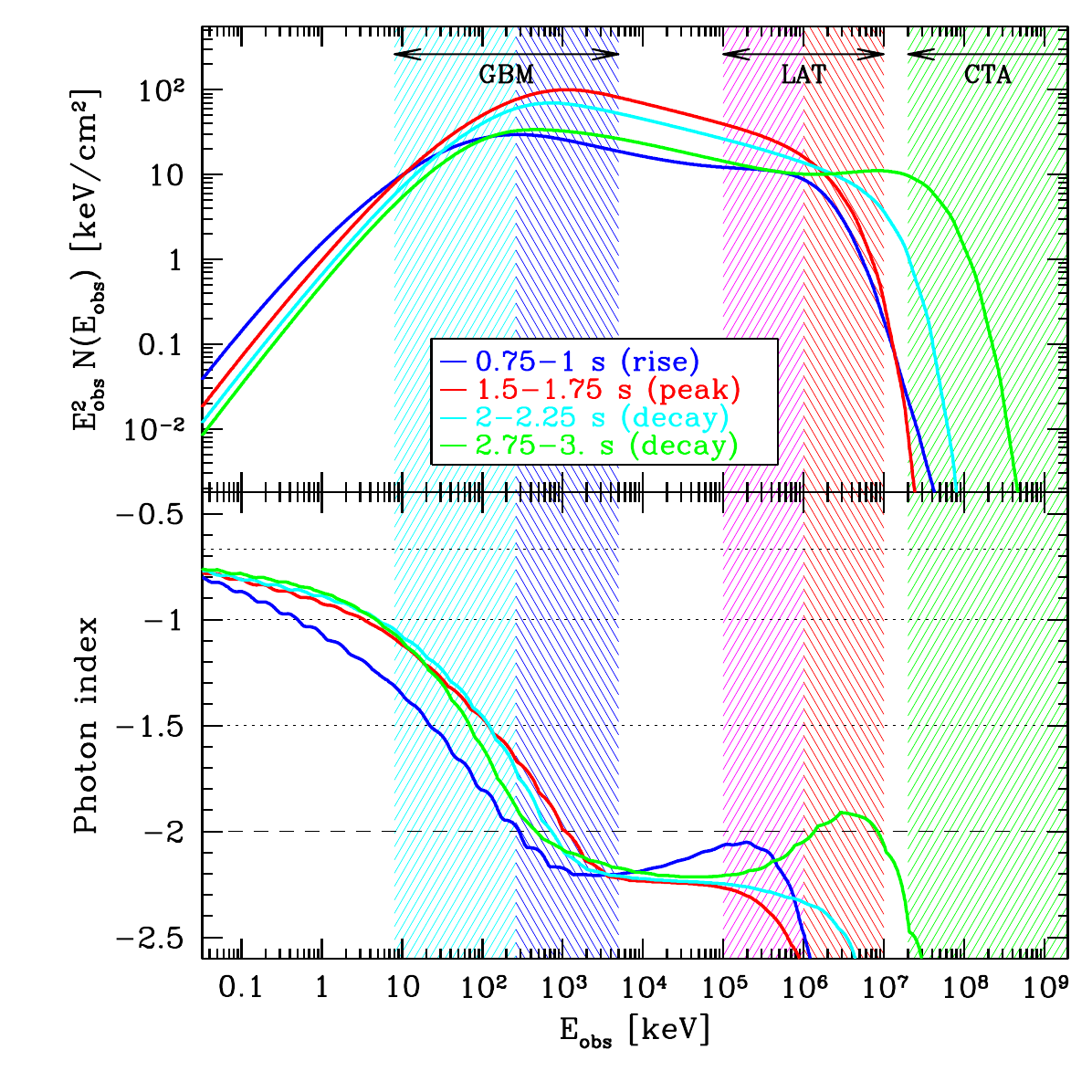}\\
    \end{tabular}
    \caption{\textbf{Effect of a decaying magnetic field in the internal shock model: reference case A with $t'_\mathrm{B}/t'_\mathrm{dyn}=10^{-4}$.}
    Left: lightcurves in the GBM and LAT range. The top panel shows the evolution of the parameters $w_\mathrm{m}$, $Y_\mathrm{Th}$ and $\Gamma_\mathrm{c,0}/\Gamma_\mathrm{m}$ in the comoving frame of the shocked material. Right: spectra in the four  time bins indicated on the lightcurves and corresponding to the rise, the peak and the decay of the pulse.
}
    \label{fig:A4m}
\end{figure*}

\begin{figure*}
    \centering
    \begin{tabular}{cc}
    \includegraphics[width=0.48\linewidth]{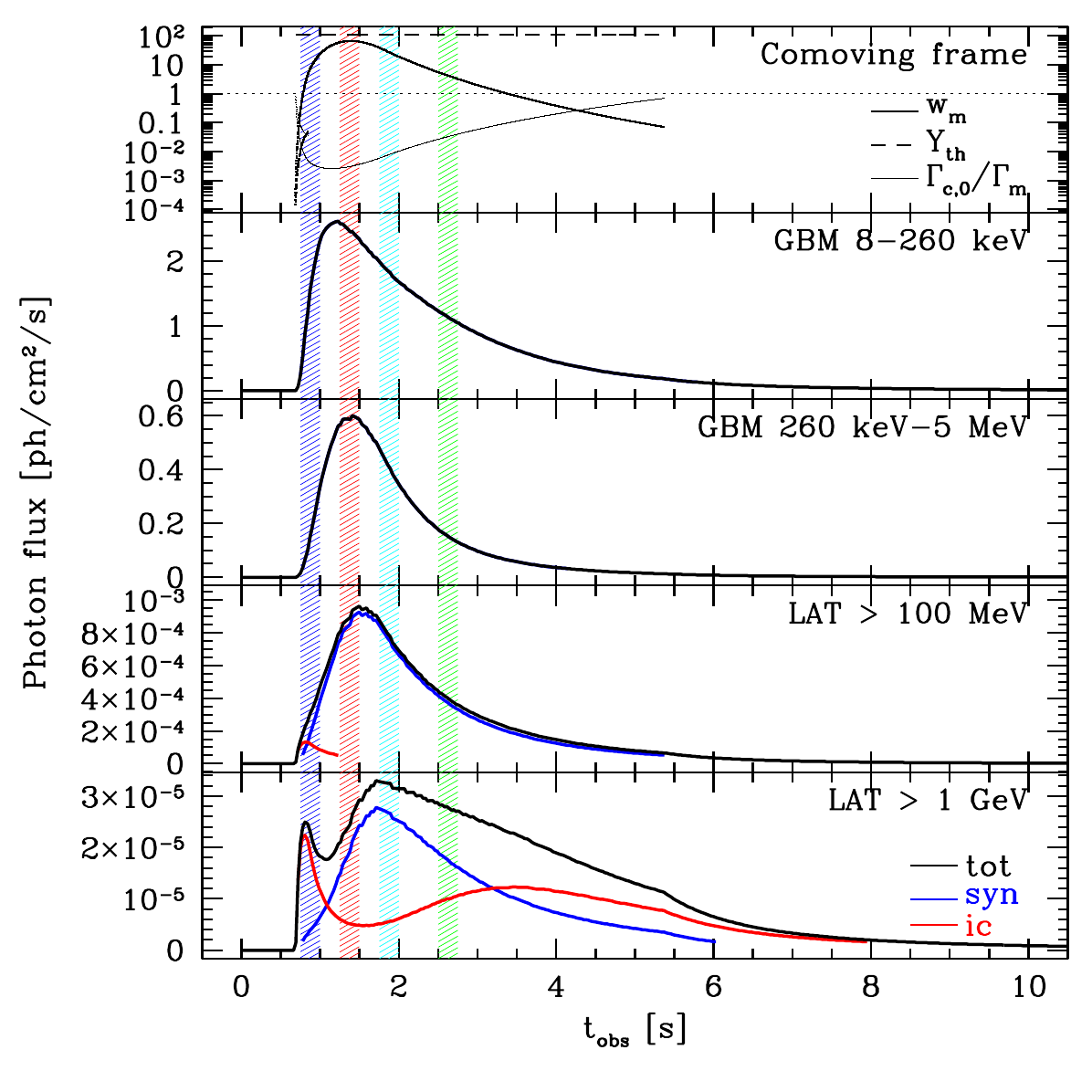} &
    \includegraphics[width=0.48\linewidth]{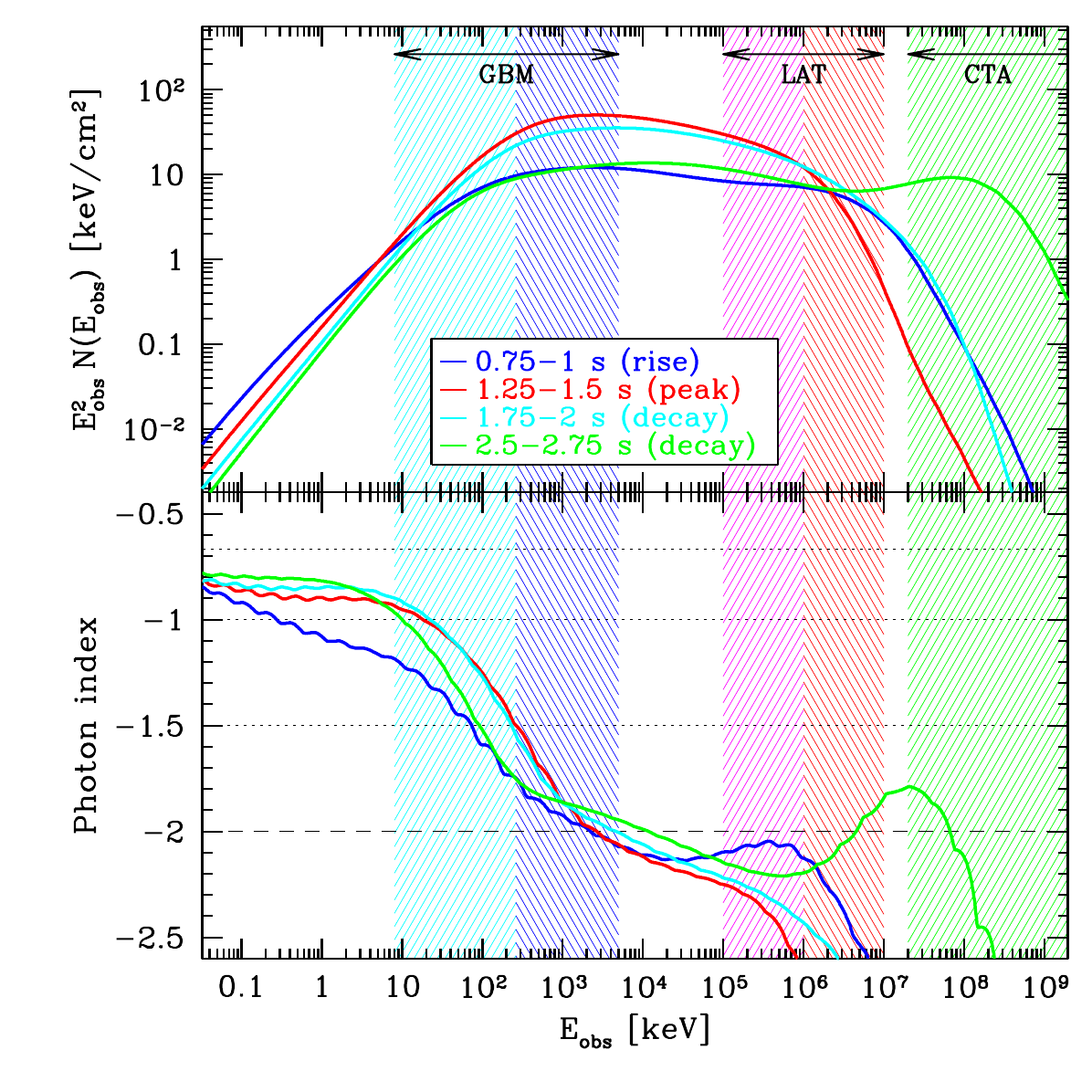}\\
    \end{tabular}
    \caption{\textbf{Effect of a decaying magnetic field in the internal shock model: reference case B with $t'_\mathrm{B}/t'_\mathrm{dyn}=10^{-3}$.}
    Same as in figure~\ref{fig:A4m}.
}
    \label{fig:B3m}
\end{figure*}

We show in \reffig{fig:A4m} and \reffig{fig:B3m} the same cases, where the effect of a decaying magnetic field is now included, assuming $t'_\mathrm{B}/t'_\mathrm{dyn}=10^{-4}$ in case (A) and $10^{-3}$ in case (B). Lightcurves in different energy channels (covering the spectral range of the \textit{Fermi}/GBM and LAT instruments) are plotted in the left panel and the time-evolving spectrum is plotted in the right panel. In addition, the evolution of the parameters governing the synchrotron and IC regime, namely $\Gamma_\mathrm{c,0}/\Gamma_\mathrm{m}$, $Y_\mathrm{Th}$ and $w_\mathrm{m}$ are plotted on the top-left panel, and the photon index is plotted in the bottom-right panel. With our choice of timescales for the magnetic field decay, the condition given by \refeq{eq:conditiondecay} is fulfilled in both cases at the maximum of the pulse.
As expected, the low-energy $\nu F_{\nu}$ 
spectrum is much steeper when including the magnetic field decay, with 
$\alpha \simeq -0.8$ 
in both cases.

We observe in addition a clear spectral evolution during the pulse, with both the peak energy and the low-energy photon index evolving from the rise to the peak of the pulse, and then from the peak to the decay, with a general hard-to-soft trend. Finally, the inverse Compton emission becomes efficient even in case (A) during the decay phase, leading in both cases to a slightly delayed emission in the GeV range (see bottom-left panel in \reffig{fig:A4m} and \reffig{fig:B3m}).

We note that all microphysics parameters ($\epsilon_\mathrm{e}$, $\zeta$, $p$, $\epsilon_\mathrm{B}$) are assumed here to be constant. Any evolution with the shock conditions would affect the details of observed spectral evolution, as discussed in \citet{daigne03,bosnjak14}. In addition, the ratio $t'_\mathrm{B}/t'_\mathrm{dyn}$ has also been assumed to be constant for simplicity. We discuss this assumption in the next subsection.

\subsubsection{Evolution of the timescale of the magnetic field decay}

\begin{figure*}
    \centering
    \begin{tabular}{cc}
    \includegraphics[width=0.45\linewidth]{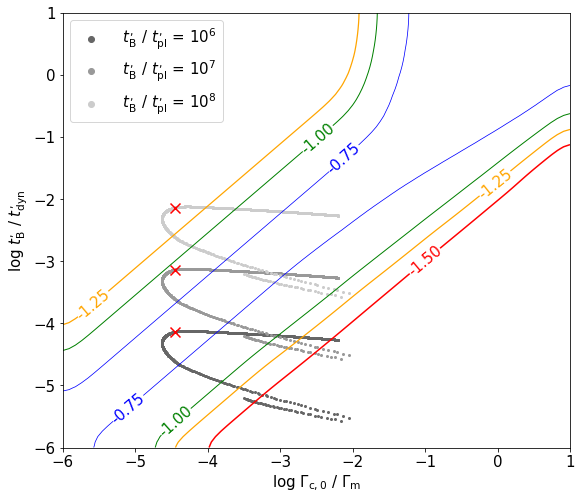} &
    \includegraphics[width=0.45\linewidth]{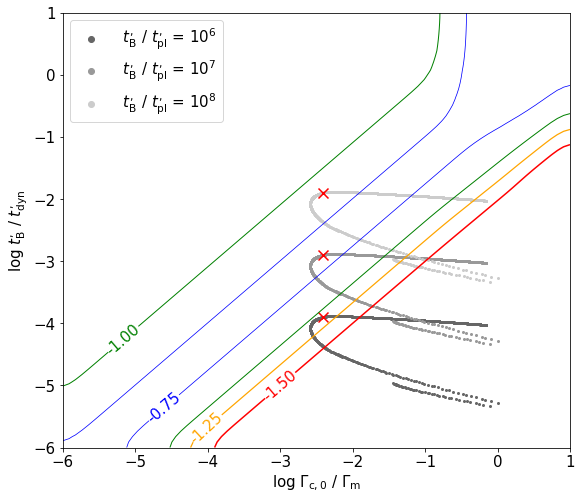}\\
    \end{tabular}
    \caption{\textbf{Evolution of $t'_B/t'_{dyn}$ and $\Gamma_{c,0}/\Gamma_m$ during the pulse duration.} {\it Left:} The evolution for {\it case A}.  The contour lines for values of $\alpha = -1.5$, $-1.25$, $-1$, $-0.75$ are shown, using the simulations obtained for $Y_{\mathrm{Th,0}} = 0.1$; $w_\mathrm{m} = 10^2$ (adopted from the upper right plot on Fig. 3.). {\it Right:} The evolution for {\it case B}. The same contour lines are shown for $\alpha$, using the simulations obtained for $Y_{\mathrm{Th,0}} = 10^2$ ; $w_\mathrm{m} = 10^2$ (adopted from the middle left plot on Fig. 3.). We overplotted (in gray lines) the values of 
$t_{\mathrm{B}}'/t_{\mathrm{dyn}}'$
    calculated using the parameters of the dynamical evolution ($\Gamma_m$ and density in the comoving frame) for {\it case A} and 
    {\it case B}
    and assuming a constant ratio $t'_\mathrm{B}/t'_\mathrm{pl}=10^6$, $10^7$ or $10^8$.
Red crosses show values at the maximum of the pulse, i.e. where the dissipated energy per unit mass obtains maximum.} 
    \label{fig:plasmadepth}
\end{figure*}

In the fast cooling regime, typical electrons responsible for the peak and the low-energy part of the synchrotron spectrum cool and radiate on a timescale which is orders of magnitude above the plasma scale at their acceleration site, but orders of magnitude below the dynamical timescale of the ejecta. Then, they probe a magnetic field on a scale which is neither accessible to particle-in-cell simulations of the acceleration process (typically $\sim 10^4$ plasma scales at maximum, see e.g. \citealt{keshet09, crumley19}) nor to large-scale MHD simulation of the propagating ejecta 
\citep[e.g.][]{bromberg2016}.
Therefore we do not have predictions for the realistic magnetic field structure to take into account for the synchrotron radiations in GRBs, that could be compared to the very simple prescription used in this paper (exponential decay) and the condition given by \refeq{eq:conditiondecay} on the decay scale. 

The spectral evolution shown in \reffig{fig:A4m} and \reffig{fig:B3m} is computed assuming a constant ratio $t'_\mathrm{B}/t'_\mathrm{dyn}$, i.e. a strong correlation between the magnetic field decay timescale and the dynamical timescale of the ejecta. Another possible assumption would be to consider a strong correlation between the decay scale and the plasma scale, such as a constant ratio $t'_\mathrm{B}/t'_\mathrm{pl}$, where $c t'_\mathrm{pl}$ is the plasma electron skin depth, given by \citep[see e.g.][]{peer06}:
\begin{equation}
t'_{{\mathrm{pl}}} = \frac{1}{\omega'_{\mathrm{pl,e}}}=\sqrt{\frac{\gamma_{\mathrm{m}} m_e}{4\pi e^2 n'_{\mathrm{e}}}}
\, ,
\label{eq:tplasma}
\end{equation}
where $n'_\mathrm{e}=n'_\mathrm{p}$ is the comoving density.
In agreement with 
\refeq{eq:conditiondecay}, \citet{peer06} found that hard synchrotron spectrum in GRBs could be expected for $t'_\mathrm{B}/t'_\mathrm{pl}\sim 10^5$. 
If we assume a constant ratio $t'_\mathrm{B}/t'_\mathrm{pl}$, the spectral evolution should be affected as the ratio $t'_\mathrm{B}/t'_\mathrm{dyn}$ is now evolving.
For the two cases shown in \reffig{fig:A4m} and \reffig{fig:B3m}, we plot in 
\reffig{fig:plasmadepth} 
the evolution of the physical conditions in the emission region in the  $t'_\mathrm{B}/t'_\mathrm{dyn}$ vs $\Gamma_\mathrm{c,0}/\Gamma_\mathrm{m}$ plane relevant for 
understanding the regime of the synchrotron emission. This evolution is plotted for $t'_\mathrm{B}/t'_\mathrm{pl}=10^6$, $10^7$ and $10^8$, which are intermediate values between the estimates considered by \citet{peer06} ($\sim 10^5$) and \citet{piran05} ($\sim 10^9$). 
This high ratio confirms that we probe here the evolution of the magnetic field on a scale which is much larger than the plasma scale\footnote{In PIC simulations of electron-ion plasmas, results are usually plotted in units of the proton skin depth $t'_\mathrm{pl,p}$. The ratio is $t'_\mathrm{pl,p}/t'_\mathrm{pl}=\sqrt{\gamma_\mathrm{m,p} m_\mathrm{p}/\gamma_\mathrm{m}m_\mathrm{e}}\sim 0.1-10$ depending on the ratio of the fractions of accelerated protons and electrons, and the ratio of the fraction of the dissipated energy injected in protons and electrons. Therefore $t'_\mathrm{B}/t'_\mathrm{pl,p}$ is also very high, and larger than the typical box size and duration of a PIC simulation.} at the acceleration site. 
For comparison, lines of constant low-energy photon index $\alpha=-1.5$, $-1.25$, $-1$ and $-0.75$ are also plotted, assuming $Y_\mathrm{Th,0}=0.1$ and $w_\mathrm{m}=10^2$ in case A, and $Y_\mathrm{Th,0}=10^2$ and $w_\mathrm{m}=10^2$
in case B, which are representative of the conditions close to the peak of the pulse (see top-left panels in \reffig{fig:A4m} and \reffig{fig:B3m}). 
A red cross indicates the time where the dissipated energy per unit mass is maximum.
Compared to a constant ratio $t'_\mathrm{B}/t'_\mathrm{dyn}$, which would correspond to an horizontal line in \reffig{fig:plasmadepth}, we find that a correlation with the plasma scale leads to evolution allowing to stay for most of the pulse in the region with the hardest low-energy spectral slope. This comparison could be improved by taking into account the evolution of $w_\mathrm{m}$ (and possibly $Y_\mathrm{Th,0}$ if other microphysics parameters such as $\epsilon_\mathrm{B}$ are also varying). 
We conclude that if the prompt GRB emission is produced by synchrotron radiation, the observed spectral evolution is most probably strongly affected by the unknown structure of the magnetic field at intermediate scales, but we do not pursue further this exploration in absence of realistic prescription.


\section{Discussion and conclusions}
\label{sec:discussion_conclusion}
If the prompt GRB emission is dominated by synchrotron radiation from electrons accelerated by internal shocks or reconnection above the photosphere, they must be in fast cooling regime to explain the observed huge luminosities and the high variability. Therefore their radiative timescale is much shorter than the dynamical timescale $t'_\mathrm{dyn}$ of the ejecta, but is also much longer than the plasma scale $t'_\mathrm{pl}$ at the acceleration site. We have explored the effect on the emitted spectrum of a possible evolution of the magnetic field on such intermediate scales, which cannot currently be probed either by large scale MHD simulations, or by PIC simulations. In the absence of any available prediction for the evolution of the magnetic field at these scales, we have adopted a simple prescription with an exponential decay on a timescale $t'_\mathrm{B}$, with $t'_\mathrm{pl}\ll t'_\mathrm{B}\ll t'_\mathrm{dyn}$. Our detailed calculations, including not only the synchrotron radiation but also a detailed treatment of inverse Compton scatterings with Klein-Nishina corrections, as well as synchrotron self-absorption at low energy and pair production at high energy, show that such a decaying magnetic field has a strong impact on the emission, with a significant steepening of the low-energy part of the $\nu F_\nu$ spectrum.  These lead to three main results:

\noindent (1)~\textit{Low-energy photon index $\alpha$ in the soft gamma-ray range:} 
a possible rapid magnetic field decay is a natural mechanism to reach the marginally fast-cooling regime \citep{daigne11,beniamini:13} with $\alpha$ close to $-2/3$ while maintaining a high radiative efficiency. 
The highest values of the low energy photon index, $\alpha\simeq -2/3$, are obtained
in a broad region of the parameter space, where the magnetic field decays on a timescale intermediate between the radiative timescale of electrons at $\Gamma_\mathrm{m}$ and the dynamical timescale $t'_\mathrm{dyn}$. Precisely, our numerical results show  that the approximate condition for this marginal fast cooling regime is 
$0.1\, {\Gamma_\mathrm{c,0}}/{\Gamma_\mathrm{m}} \lesssim {t'_\mathrm{B}}/{t'_\mathrm{dyn}} \lesssim 10\, {\Gamma_\mathrm{c,0}}/{\Gamma_\mathrm{m}}$. 
This corresponds to a characteristic scale of the magnetic field decay which is typically $\sim 7-8$ orders of magnitude above the plasma skin depth at the acceleration site, in agreement with the results of previous studies \citep{piran05,peer06}.
For a slower magnetic field decay, $10\, {\Gamma_\mathrm{c,0}}/{\Gamma_\mathrm{m}} \lesssim {t'_\mathrm{B}}/{t'_\mathrm{dyn}}$, the effect of the magnetic field decay is much weaker. Then the low-energy photon index is in the range $-3/2\le \alpha \lesssim -1$, depending on the importance of inverse Compton scatterings in the Klein-Nishina regime. On the other hand, a much faster decay (${t'_\mathrm{B}}/{t'_\mathrm{dyn}}  \lesssim 0.1\, {\Gamma_\mathrm{c,0}}/{\Gamma_\mathrm{m}}$) can lead to much flatter spectra ($-2 < \alpha < -3/2$), as the whole electron population enters the slow cooling regime. This leads to a low radiative efficiency. 
A non negligible background magnetic field may however very well dominate the electron evolution in this specific regime \citep{zhou:23}.
We have shown that by varying the physical conditions in the emission region (electron Lorentz factor at injection, magnetic field, electron density, dynamical timescale), all these different regimes of the synchrotron fast-cooling can be explored. 

\noindent (2)~\textit{Spectral evolution.} During the GRB prompt emission, synchrotron radiation is expected from  accelerated  electrons in different emission zones (e.g. shocks or reconnection sites), which leads to the observed diversity and variability of GRB light curves. In addition, the physical conditions in each  emission zone can evolve, which naturally leads to some spectral evolution even within a single pulse. 
Such a spectral evolution is observed in the time-resolved spectral analysis of bright GRBs  \citep[see e.g.][]{yu2016}.
We have have shown some examples of the spectral evolution expected in the case of a pulse produced by an internal shock in Section~\ref{sec:pulse_is}: the peak energy and the low-energy photon index vary significantly during the rise and the decay of the pulse light curve. However the predicted evolution depends strongly on the assumed prescription for the characteristic scale of the magnetic decay. This prescription is very uncertain in the absence of simulations covering the full range of scales from the plasma scale at the acceleration site to the geometrical width of the ejecta. The results shown here are limited to the specific case where $t'_\mathrm{B}$ is strongly correlated to the dynamical timescale  $t'_\mathrm{dyn}$. 

\noindent (3)~\textit{High-energy emission:} 
\textit{Fermi}/LAT observations show some diversity regarding the prompt high-energy emission
\citep{ajello19}: the ratio between the fluences in the 0.1-100 GeV and 10 keV-1 MeV channels ranges \mbox{from $\sim 10^{-2}$ to $\sim 1$} for the cases with a LAT detection (during the GBM time window). The synchrotron self-Compton component in Klein-Nishina regime is a natural candidate to explain this behaviour \citep{bosnjak09}. We have show here that it is also affected by a possible
magnetic field decay in the emission zone. 
Indeed, the effective Compton parameter increases with time as the magnetic field decays. As discussed in Section~\ref{sec:results}, our exploration of the parameter space shows 
ratios of the intensity of these two components
$L_\mathrm{ic}/L_\mathrm{syn}$ ranging from very low values when the Klein-Nishina reduction is strong, to larger values of the order of $L_\mathrm{ic}/L_\mathrm{syn}\simeq 0.1-1$, in agreement with the observed diversity of the prompt high-energy component.

Our main conclusion is that \textit{efficient} synchrotron radiation in a rapidly decaying magnetic field can reproduce low-energy photons indices ranging from $\alpha=-3/2$ to $-2/3$, in agreement with the measured value in the majority of GRBs \citep[see e.g.][]{goldsteinbatse,gruber2014,poolakkil2021}. For instance, the peak spectrum of the GRBs in the best sample of the \textit{Fermi}/GBM spectral catalog shows $\alpha<-2/3$ in about 78\% of cases, with a median value and 68\% confidence level interval $\alpha=-1.30^{+0.77}_{-0.33}$ \citep{poolakkil2021}, and the time-resolved spectral analysis of the brightest \textit{Fermi}/GBM over the four first years show a median value 
$\alpha=-0.77^{+0.27}_{-0.32}$
\citep{yu2016}. We note that the measured value of $\alpha$ is affected by the imposed shape of the phenomenological function used for the spectral fit, usually the Band function \citep{band93}.  \citet{oganesyan17,ravasio2018,toffano21,poolakkil2023} obtain good spectral fits to several bright GRBs by adding a second break in the spectrum, often close to the peak energy, with low-energy photon indices close to $-2/3$ and $-3/2$, which is reminiscent of the marginally fast cooling regime naturally favored by the magnetic field decay described here. However, we also note that the theoretical synchrotron spectral shape cannot be directly compared to data, as it should be convoluted with the evolution of the physical conditions in the emission zone, which tends to smooth and broaden the spectral breaks, as illustrated in the simulations shown in Section~\ref{sec:pulse_is} in the context of the internal shock model. \citet{yassine2020} introduced the ISSM phenomenological function with four free parameters, which has a smoothly evolving photon index all along the soft gamma-ray range and fits well synthetic spectra produced by fast cooling synchrotron radiation in internal shocks. They find that the ISSM can usually fit at least as well than the Band function the spectra of a sample of 74 bright \textit{Fermi/GBM} bursts, with a low-energy photon index $\alpha<-2/3$ in 74\% of cases. Finally we note that in all scenarios for the prompt emission where soft gamma-rays are produced by synchrotron radiation in the optically thin region, a weak quasi-thermal photospheric emission is also expected \citep[see e.g.][]{hascoet13}. 
\citet{guiriec2015,guiriec2017,li2019} have shown that adding a weak thermal component in the spectral fit can also significantly affect the measured value of the low-energy photon index $\alpha$ of the non-thermal component, usually by decreasing it. For a better comparison, it would be necessary in the future to be able to perform the spectral analysis of GRB data using directly a detailed synchrotron model as described in this paper, coupled to a dynamical calculation of the relativistic ejecta, and probably also including the contribution of the photospheric emission. This goes far beyond the scope of the present paper, and seems difficult to achieve in the short term. On the other hand, there are 
good short-term prospects for a better measurement of the prompt GRB spectrum, and especially the low-energy index $\alpha$, thanks to the SVOM mission to be launched in June 2024 \citep{wei:16}, using the low-energy threshold at 4 keV of the ECLAIRs coded-mask telescope combined by the spectral range extending to 5 MeV of the GRM detectors  \citep{bernardini:17}.

 
\begin{acknowledgements}
The authors thank Robert Mochkovitch, Frédéric Piron and Martin Lemoine for fruitful discussions about this study.
The research leading to these results has received funding (\v Z.B.) from the European Union’s Horizon 2020 Programme under the AHEAD2020 project (grant agreement n. 871158). 
F.D. acknowledges the Centre National d’Études Spatiales (CNES) for financial support in this research project.
      
\end{acknowledgements}

\bibliographystyle{aa}
\bibliography{biblio.bib} 

\end{document}